\documentclass[preprint,showpacs,preprintnumbers,amsmath,amssymb]{revtex4-1}
\usepackage{graphicx}
\usepackage{color}
\usepackage{bm}
\usepackage{amsmath}
\usepackage{amssymb}
\usepackage{setspace}
\usepackage{array}
\usepackage{tabularx}
\usepackage{multirow}
\usepackage{afterpage}
\usepackage{hhline}
\usepackage{url}
\usepackage{hyperref} 
\usepackage{dcolumn}
\usepackage{footmisc} 
\usepackage{xspace}
\usepackage{natbib}
\newcolumntype{P}[1]{>{\centering\arraybackslash}p{#1}}
\newcolumntype{M}[1]{>{\centering\arraybackslash}m{#1}}
\begin{document}

\title{Strongly Interacting Ultracold Polar Molecules}

\author{Bryce Gadway}
 \affiliation{Department of Physics, University of Illinois at Urbana-Champaign, Urbana, IL 61801, USA}
  \email{bgadway@illinois.edu}
\author{Bo Yan}%
 \affiliation{Department of Physics, Zhejiang University, Hangzhou, Zhejiang, 310027, China}%
  \email{yanbohang@zju.edu.cn}

\date{\today}

\begin{abstract}
This paper reviews recent advances in the study of strongly interacting systems of dipolar molecules. Heteronuclear molecules feature large and tunable electric dipole moments, which give rise to long-range and anisotropic dipole-dipole interactions. Ultracold samples of dipolar molecules with long-range interactions offer a unique platform for quantum simulations and the study of correlated many-body physics. We provide an introduction to the physics of dipolar quantum gases, both electric and magnetic, and summarize the multipronged efforts to bring dipolar molecules into the quantum regime. We discuss in detail the recent experimental progress in realizing and studying strongly interacting systems of polar molecules trapped in optical lattices, with particular emphasis on the study of interacting spin systems and non-equilibrium quantum magnetism. Finally, we conclude with a brief discussion of the future prospects for studies of strongly interacting dipolar molecules.
\end{abstract}

\singlespacing %
\maketitle %

\tableofcontents

%
\section{Introduction}
\label{sec:sec1intro}

We are at a very exciting time in the study of dipolar quantum gases. Gases of atoms and molecules with large magnetic and electric dipole moments have been in the quantum gas realm for nearly a decade, since the realization of a Bose--Einstein condensate of chromium ($^{52}$Cr) atoms in 2005 by the Pfau group in Stuttgart~\cite{ChromiumBEC-2005}, followed several years later by the demonstration of a nearly degenerate Fermi gas of heteronuclear potassium-rubidium ($^{40}$K$^{87}$Rb) molecules by the Jin and Ye collaboration at JILA~\cite{ni2008}. The intervening years have seen many important efforts, both experimental and theoretical, to observe and understand the role of dipole-dipole interactions in ultracold quantum gases. These include the study of equilibrium properties and dynamics~\cite{Stuhler:Cr_BEC_TOF2005,lahaye:strong_chromium_2007,Lahaye:Cr_collappse2008,Bismut:Cr_collecive2010,ni2010,Bismut-ExcSpec-2012,lu:strongly_Dy_2011,aikawa:bose-einstein_Er_2012,Aikawa:Er_Fermi_surface2014}, analyses of stabilization against mean-field collapse or chemical reactions~\cite{Koch-Stability,silke2010,LatticeStability-Muller,miranda2011,chotia:long-lived_2012}, and novel phenomena arising due to the dipolar coupling of spin and motional degrees of freedom~\cite{Fattori-DemagCooling,Pasquiou-SpinRelax,QHE-dipolar-Pfau,Ueda-EDH}, just to name a few.

\begin{figure}[b]
\includegraphics[height=8cm]{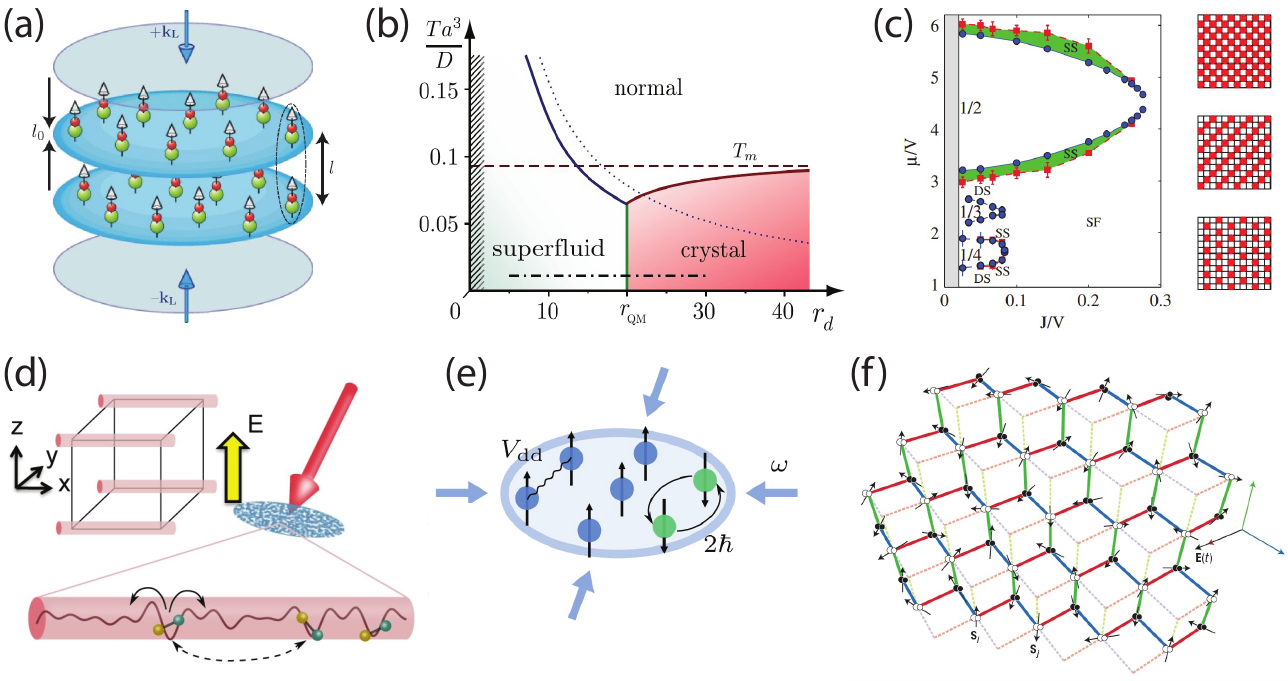}
\caption{\label{Fig:TheoryPlaquette}
Menagerie of some selected novel phenomena proposed to be studied with long-ranged interacting dipolar atoms and molecules. (\textbf{a})~Fermionic dipoles in a layered planar geometry can undergo novel pairing mechanisms. Reprinted figure with permission from~\cite{Baranov-Bilayer-image}. Copyright (2011) by the American Physical Society. (\textbf{b},\textbf{c})~The respective phase diagrams for planar fermionic dipoles exhibiting a Wigner crystal-like phase and planar bosonic lattice dipoles exhibiting supersolid (SS) and charge density wave ordering, respectively. Reprinted figures with permission from~\cite{Buchler-Crystal-2007} and ~\cite{Sansone-2Dphases-2010}. Copyrights (2007) and (2010) by the American Physical Society. (\textbf{d})~The interplay of disorder and long-ranged dipolar interactions can give rise to many-body localized phases. Reprinted figure with permission from~\cite{Yao2014:localization}. Copyright (2014) by the American Physical Society. (\textbf{e})~Dipolar coupling of internal (spin) and external (orbital) angular momentum can be used to study nontrivial topological states. Reprinted figure with permission from~\cite{QHE-dipolar-Pfau}. Copyright (2013) by the American Physical Society. (\textbf{f})~Localized molecules interacting via dipole-dipole interactions, simulating an interacting spin model. Reprinted with permission from Macmillan Publishers Ltd: Nature Physics~\cite{Micheli_NP_2006}, copyright~(2006).
}
\end{figure}

In just the past few years, researchers have begun to observe the dynamics of strongly interacting systems driven purely by long-range dipolar interactions in experiments with polar molecules~\cite{yan2013:dipole-dipole,Hazzard2014} and with magnetic atoms~\cite{dePaz-Chrom-QM-2013}. Importantly, the basic dipolar processes underlying these observations are integral to a bevy of proposals for the realization of exotic states of matter and the study of non-trivial dynamics in dipolar systems. As highlighted by the selection of images in Fig.~\ref{Fig:TheoryPlaquette}, such proposals address the study of novel paired superfluid phases~\cite{Baranov-Bilayer-image,potter:2006,Baranov-DipolePairing}, dipolar crystals~\cite{Buchler-Crystal-2007} and supersolids~\cite{Scarola-SSphase-2006,Menotti-SSphase-2007,Sansone-2Dphases-2010,Buchler-SS}, disorder and many-body localized phases~\cite{Yao2014:localization}, topological fluids~\cite{QHE-dipolar-Pfau,Syzranov2014:SOC}, and myriad other phenomena. Especially promising is the ability of dipolar atoms and molecules to simulate quantum magnets~\cite{barnett:quantum_2006,gorshkov} and interacting spin systems~\cite{Micheli_NP_2006}, as coherent spin dynamics have been shown to be extremely robust even at high entropies when particle motion is quenched in a deep optical lattice~\cite{yan2013:dipole-dipole,Hazzard2014}.

The coming years promise to be a time of great progress in the study of dipolar systems, as researchers move beyond initial observations of interaction-driven dynamics to study complex phenomena in regimes that are theoretically intractable by currently available techniques. With many new research groups and species of atom and molecule~\cite{Takekoshi:STIRAP2014,Molony:STIRAP2014,Park-NaK-GS-2015,NaRbGas} joining the effort, dipolar quantum gas research is poised to expand greatly in both breadth and capability. At the same time, the study of dipolar spin physics in quantum gas experiments will benefit from vigorous competition from other atomic, molecular, and optical (AMO) platforms such as Rydberg atoms~\cite{Pfau-Forster-12,Browaeys-DipDip-14,Browaeys-ExcitationTransfer-15}, Rydberg-dressed atoms~\cite{RydDress-Pohl-2010,RydDress-Pupillo-2010,RydDress,RydDressLattice}, and ions with effective dipolar interactions~\cite{Wilson-DipDip-Ions-2014}, as well as solid state platforms such as nitrogen-vacancy (NV) centers in diamond~\cite{dolde-nv-2013,Plenio-simulator-2013}. At this juncture, we seek to present a topical review of the recent advances in the study of strongly-interacting dipolar gases, with particular emphasis on experiments involving cold polar molecules trapped in optical lattices. Here, we primarily focus on aspects of molecules gases related to quantum simulation studies, whereas a more thorough discussion of the formation of ultracold molecules may be found in several other reviews~\cite{Review-EurPhysJD,carr:cold_2009,Review-Dulieu-RepProg,Review-PCCP-2011,Quemener2012:control} and books~\cite{BookKrems} (including the 2006 Special Issue of \emph{J. Phys. B}~\cite{Review-Dulieu-JPB2006}).

This review is organized as follows. We begin with a basic introduction into the properties of dipole-dipole interactions and a discussion on dipolar quantum gases. We especially emphasize the physics of electrically dipolar molecules, and discuss efforts being made to bring a diverse group of molecular species to the quantum gas regime. We then discuss in detail systems of strongly interacting polar molecules in optical lattices. We discuss how polar molecules can be used to study long-ranged interacting spin models, and we provide a discussion of experimentally relevant aspects of such quantum simulation studies, such as the mitigation of particle loss and spin dephasing. Next, we describe in detail the experimental observation of coherent many-body spin dynamics in a system of lattice-confined polar molecules. Finally, we conclude with a discussion of the prospects for quantum simulations involving strongly interacting dipolar systems.

\section{Dipolar gases}
\label{sec:sec2dipolar}

We begin by describing the basic properties of dipole-dipole interactions and dipolar quantum gases. We then describe in more detail some particular aspects of quantum gases of magnetically dipolar atoms and electrically dipolar molecules, contrasting some of their specific attributes for quantum simulation studies. Lastly, focusing on systems of molecules, we briefly review the suite of techniques being developed to bring new molecular species into the quantum realm.

\subsection{Dipole-dipole interactions}

\begin{figure}
\includegraphics[height=6cm]{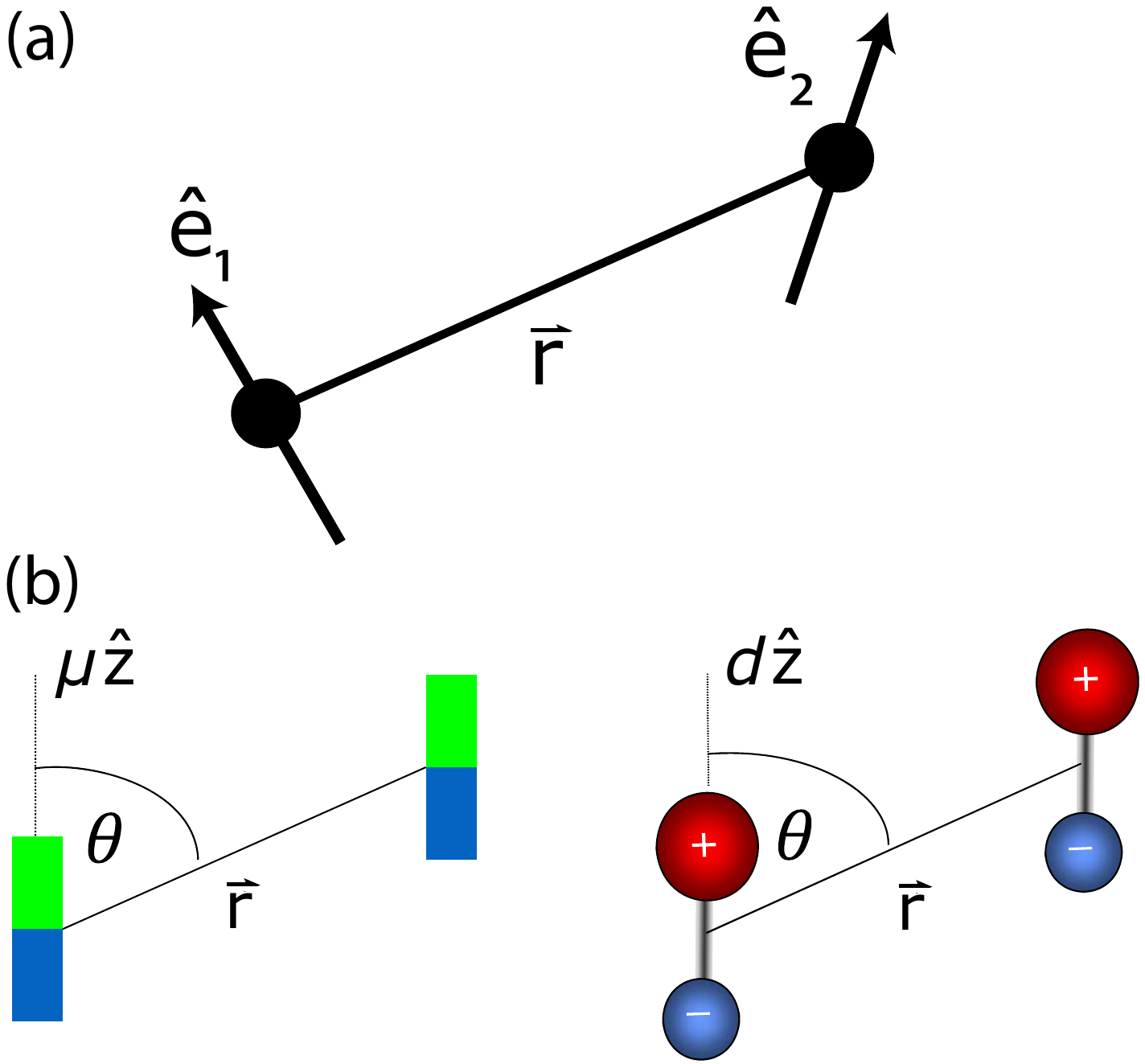}
\caption{\label{Fig:two_dipoles}
Dipole-dipole interactions. (\textbf{a})~For the general case of two particles, having their respective dipole moments pointed along the unit vectors $\hat{e}_1$ and $\hat{e}_2$ and with relative position described by the separation vector $\vec{r}\equiv r \hat{r}$, the potential energy due to dipole-dipole interactions is described by Eq.~\ref{equ:dipdip1}. (\textbf{b})~Dipole-dipole interactions between two dipoles pointed along the same direction (along unit vector $\hat{z}$). In this case the dipole-dipole interaction is described by a simplified form $V_{dd}(\vec{r})= C_{dd} (1 - 3\cos^2\theta)/ 4\pi r^3$, where $\theta$ is the angle between the dipole moments and the particles' relative position vector $\vec{r}$. For the case of magnetic dipoles, as shown at left, the interaction coefficient is given as $C_{dd} = \mu_0 \mu^2$ (where $\mu_0$ is the permeability of free space and $\vec{\mu} \equiv \mu \hat{z}$ is the magnetic dipole moment), and for electric dipoles, as shown at right, the interaction coefficient is given as $C_{dd} = d^2/\epsilon_0$ (where $\epsilon_0$ is the permittivity of free space and $\vec{d} \equiv d \hat{z}$ is the electric dipole moment).
}
\end{figure}

We consider the general case of two interacting particles (labeled 1 and 2 as shown in Fig.~\ref{Fig:two_dipoles}~(a)), having relative position $\vec{r} \equiv r \hat{r}$ and with dipole moments pointed along the unit vectors $\hat{e}_1$ and $\hat{e}_2$, respectively. The potential energy of the two particles due to their dipole-dipole interactions (DDIs) is given by
\begin{equation}
V_{dd}(\vec{r})=\frac{C_{dd}}{4\pi}\frac{\hat{e}_1\cdot \hat{e}_2 - 3(\hat{e}_1\cdot \hat{r})(\hat{e}_2\cdot \hat{r})}{r^3}.
\label{equ:dipdip1}
\end{equation}
This description applies for both magnetic and electric dipoles, where the coupling constant $C_{dd}$ is given by $\mu_0 \mu^2$ for particles with a permanent magnetic dipole moment of magnitude $|\vec{\mu}| = \mu$ (where $\mu_0$ is the permeability of free space) and by $d^2/\epsilon_0$ for particles having permanent electric dipole moments with common magnitude $|\vec{d}| = d$ (where $\epsilon_0$ is the permittivity of free space).

When these dipoles are fixed to point along a common direction $\hat{z}$, as in a sample spin-polarized with respect to a laboratory quantization axis, this expression is simplified to the form
\begin{equation}
V_{dd}(\vec{r})=\frac{C_{dd}}{4\pi}\frac{1 - 3\cos^2\theta}{r^3},
\end{equation}
where $\theta$ is the angle of intersection between the lab-frame dipole moments and the particles' relative position vector, as shown in Fig.~\ref{Fig:two_dipoles}~(b). Two defining characteristics of dipole-dipole interactions are their natural anisotropy (dependence on $\theta$) and the fact that they are relatively long-ranged (scaling with distance in free space as $1/r^3$). These features help to make dipolar particles extremely interesting from a quantum simulation standpoint, as they naturally give rise to a number of interesting interaction-driven effects not present (at lowest order) in systems with purely short-ranged contact interactions. In addition to the described static part of the dipole-dipole interaction, dipolar interactions also play host to processes that can change the dipoles' orientations, in ways that either conserve or do not conserve the net magnetization (spin polarization) of the sample. The magnetization non-conserving dipolar interactions involve exchange of the particles' internal (spin) and external (motional) angular momentum and energy, and lead to dipolar relaxation~\cite{Pfau-theory-2003}, allow for novel adiabatic demagnetization-based cooling schemes~\cite{Fattori-DemagCooling}, and underly a number of proposals for the study of topological properties of quantum gases~\cite{QHE-dipolar-Pfau,Syzranov2014:SOC}. Moreover, the magnetization conserving processes underly the herein described studies of nonequilibrium quantum magnetism performed with strongly interacting polar molecules, and will be discussed further in Sec.~\ref{sec:sec3exp}.

In the context of bulk dipolar quantum gas experiments, it is instructive to characterize the degree to which dipolar interactions will influence the state of the system, as compared to either short-ranged contact interactions for bosons or to the role of Pauli exclusion for fermions, for example. In the former case, the interactions between two atoms due to $s$-wave collisions can be well described by a delta function pseudo-potential of the form
\begin{equation}
U(\vec{r})= \bar{U} \delta(\vec{r}),
\end{equation}
where the parameter $\bar{U} \equiv 4 \pi \hbar^2 a / m$ describes the influence of two-body contact collisions, $m$ is the particle mass, and $a$ is the $s$-wave scattering length. The mean-field interaction energy due to $s$-wave collisions $n \bar{U}$, where $n$ is the particle density, can thus be compared to the corresponding quantity for dipolar interactions, $n C_{dd}$. The ratio
\begin{equation}
\epsilon_{dd} = C_{dd}/3\bar{U}
\end{equation}
has therefore been used to characterize the dipolar nature of such a system (with the factor of 3 included such that a value $\epsilon_{dd}<1$ relates to stability against dipolar collapse in the case of a homogeneous dipolar Bose--Einstein condensate in three dimensions~\cite{lahaye:dipolar_review_2009}). Furthermore, an effective ``dipolar length'' $a_{dd}$ can be used to characterize dipole-dipole interactions in a fashion similar to the use of $s$-wave scattering lengths for contact interactions, defined as
\begin{equation}
a_{dd}= \frac{C_{dd} m}{12 \pi \hbar^2}.
\end{equation}

For a spin-polarized Fermi gas of dipolar particles, a natural energy scale to which one may compare $n C_{dd}$ (with $n$ the uniform particle density of the single spin state) is the Fermi energy
\begin{equation}
E_F = \frac{\hbar^2}{2m}(6\pi^2 n)^{2/3} ,
\end{equation}
thus defining the ratio
\begin{equation}
\eta_{dd}= \frac{n C_{dd}}{E_F} = \frac{4}{\pi}k_F a_{dd}
\end{equation}
to characterize the dipolar nature of a single-component Fermi gas~\cite{Aikawa:Er_Fermi_surface2014}. Here, we have also expressed this in terms of the Fermi wavevector $k_F \equiv \sqrt{2 m E_F}/\hbar$ multiplied by the dipolar length, revealing that a fermionic gas becomes more dipolar for higher particle densities and correspondingly larger Fermi energies and Fermi momenta. This can equivalently be seen as comparing the dipolar length to the typical interparticle spacing $r_0$ (taken as $r_0 = n^{-1/3} = (6\pi)^{1/3}/k_F$), where dipolar lengths on the order of $r_0$ correspond to values of $\eta_{dd} \sim 4 (6/\pi)^{1/3} \gtrsim \pi$.

In Table~\ref{tab:title}, we compare some of the characteristic dipolar parameters for several species of atom and molecule being pursued in experiment, along with some of the most polar molecules.

talk a little bit about the choices -- in particular cite the works here. We enumerate the dipole moment of the particle, the dipolar length $a_{dd}$, as well the potential energy $V_{nn}$ of two point-like particles separated by $532$~nm, relating to a nearest-neighbor lattice spacing commonly used in experiment. A cursory investigation of Table~\ref{tab:title} reveals that atoms with magnetic dipole-dipole interactions (MDDIs) behave as much weaker dipoles in comparison to molecules with electric dipole-dipole interactions (EDDIs). This disparity can be motivated by comparing simple estimates for the typical dipolar interaction energies in these two cases. The typical magnetic dipole moment for atoms is on the order of the electron magnetic moment, i.e. approximately one Bohr magneton $\mu_B = e \hbar / 2 m_e = h \times 1.39962 \ \mathrm{MHz/G}$ ($e$ and $m_e$ the electron charge and mass), ranging from roughly 1~$\mu_B$ for alkali atoms to the largest value of 10~$\mu_B$ for the lanthanoid atoms dysprosium~\cite{lu:strongly_Dy_2011,lu:quantum_Dy_2012,Pfau-Crystal-2015} and terbium. For polar molecules, an order-of-magnitude estimate of $d \sim e a_0 / 2 \approx 1.27$~D (where 1~Debye, 1~D~$= 3.336 \times 10^{-30}$~C~m) can be made by assuming a net electron charge separated by a distance of the Bohr radius $a_0 = \hbar / m_e c \alpha \approx 5.29 \times 10^{-11}$~m (with $c$ the speed of light in vacuum and $\alpha$ the fine structure constant). Comparing the respective dipolar interaction energies $C_{dd}$ for these systems yields a ratio of
\begin{equation}
\frac{\mu_0\mu_B^2}{(e a_0)^2/4\epsilon_0} = \alpha^2 \sim 10^{-4} ,
\end{equation}
which is consistent with the trends shown in Table~\ref{tab:title}.

\begin{table}[b]
\begin{tabular}{|l|l|l|l|}
\hline
 particle & dipole moment & $a_{dd} \ [a_0]$ & $V_{nn}/h [\mathrm{Hz}]$ \\
 \hline
 \hline
$^{87}$Rb & $0.5~\mu_B$  &   $0.18$ & 0.02\\
$^{52}$Cr & $6.0~\mu_B$  &  $15$  & $3.1$\\
$^{164}$Dy       &  $9.9~\mu_B$ &  $130$ & $8.4$ \\
$^{168}$Er     &    $7.0~\mu_B$  &  $67$ & $4.2$  \\
$^{168}$Er$_2$ & $14.0~\mu_B$ & $533$ & $16.9$ \\
\hline
KRb  & 0.57~D & $3.9\times10^3$ & $0.32 \times 10^3$\\
RbCs &  1.2~D &  $3.0\times 10^4$  & $1.4 \times 10^3$\\
NaK  & 2.7~D & $4.4\times10^4$ & $7.3 \times 10^3$\\
NaRb  & 3.3~D & $1.1\times10^5$ & $1.1 \times 10^4$\\
\hline
KCs  & 1.9~D & $6.0\times10^4$ & $3.7 \times 10^3$\\
LiK  & 3.5~D & $5.3\times10^4$ & $1.2 \times 10^4$\\
LiRb  & 4.1~D & $2.0\times10^5$ & $1.7 \times 10^4$\\
LiCs  & 5.5~D & $4.0\times10^5$ & $3.0 \times 10^4$\\
RbSr  & 1.5~D & $3.9\times10^4$ & $2.4 \times 10^3$\\
RbYb  & 0.21~D & $1.1\times10^3$ & 44 \\
LiYb  & $\lesssim$~0.1~D & $\lesssim \ 200$ & $\lesssim \ 10$ \\
\hline
CaF &   3.1~D  &  $ 5.3\times10^4$  & $9.6 \times 10^3$ \\
BaF &   3.2~D  &  $ 1.5\times10^5$  & $1.0 \times 10^4$ \\
SrF  &  3.5~D   &  $1.2\times10^5$ & $1.2 \times 10^4$ \\
YO  &   6.1~D  &  $3.7\times10^5$ & $3.7 \times 10^4$ \\
\hline
$^*$OH  &  1.7~D   &  $4.5\times10^3$ & $2.8 \times 10^3$ \\
CH$_3$F  &  1.9~D   &  $1.1\times10^4$ & $3.4 \times 10^3$ \\
H$_2$CO  &  2.3~D   &  $1.5\times10^4$ & $5.4 \times 10^3$ \\
ND$_3$ &   1.5~D  &  $ 4.1\times10^3$  & $2.2 \times 10^3$ \\
\hline
SrO  &   8.9~D  &  $7.7\times10^5$ & $7.9 \times 10^4$ \\
KBr &   10.4~D  &  $ 1.2\times10^6$  & $1.1 \times 10^5$ \\
\hline
\end{tabular}
\caption{Characteristic quantities relating to various dipolar atoms and molecules. The second column lists the dipole moments, given in units of $\mu_B$ for magnetic dipoles and $\mathrm{D}$ for electric dipoles. The third column shows the dipolar length $a_{dd}$ (in units of $a_0$), directly comparable across species. The last column lists the characteristic nearest-neighbor energy shift $V_{nn} = C_{dd}/4\pi (\lambda/2)^3$ for point-like particles separated by one lattice spacing in a lattice formed by laser light of wavelength $\lambda = 1064$~nm. The parameter values for $^{168}$Er$_2$ are based simply on the expected maximum magnetic dipole moment being twice that of the $^{168}$Er atom, and not on recently measured experimental values~\cite{Frisch-Mols-2015}. The listed permanent electric dipole moment for LiYb is based on Refs.~\cite{LiYbSmall1,LiYbSmall2}, while a slightly larger value was reported in Ref.~\cite{LiYbBig1} (and much larger dipole moments are generally expected for molecules incorporating metastable electronic states~\cite{Gupta-GroundExcited}). The parameter values for $^{87}$Rb relate to atoms in the state $|F,m_F\rangle = |1,1\rangle$ ground state, given as absolute magnitudes.} \label{tab:title}
\end{table}

While the dipolar interactions are typically orders of magnitude weaker for magnetic atoms, many seminal features of dipolar quantum fluids have so far only been observed in these systems~\cite{lahaye:strong_chromium_2007,Aikawa:Er_Fermi_surface2014}. This partly results from the fact that the suite of capabilities for cooling and manipulating ultracold atoms has not yet been fully extended to the study of ultracold molecules. To a large degree, however, this is virtue of the previously discussed fact that it is not necessarily the \emph{absolute} energy scale of dipolar interactions that matters, but rather the \emph{relative} contribution of dipolar interactions to the total energy.

The other typical energy scales of relevance include the thermal kinetic energy $k_B T$ ($k_B$ the Boltmann constant and $T$ the temperature), the coherent tunneling energy $t$ in an optical lattice environment, interaction energies due to $s$-wave collisions, and for fermions the Fermi energy $E_F$. In systems of magnetically dipolar bosonic atoms, Feshbach resonances~\cite{Chin2010} have been used to tune the $s$-wave scattering length $a_s$ to be very small in magnitude, thus enhancing the relative effect of dipolar interactions~\cite{lahaye:strong_chromium_2007}, and even allowing the observation of spontaneously ordered droplet structures in dipolar quantum gases of dysprosium~\cite{Pfau-Crystal-2015}. In addition to the control of short-range interaction energies, motional energy scales can also be widely tuned by use of an optical lattice. The tunneling energy and the temperature of particles can be tuned over a large range by scaling the effective particle mass, while the long-ranged interactions due to dipole-dipole coupling are relatively unaffected (in contrast to interactions driven by virtual second-order tunneling~\cite{trotzky:time-resolved_2008}). This powerful control has been used with both magnetic atoms~\cite{dePaz-Chrom-QM-2013,bruno-nonequ-2-2015,Ferlaino-Extended-2015} and with electrically dipolar molecules~\cite{yan2013:dipole-dipole,Hazzard2014} to reduce the motional energy scales of particles in the ground band, allowing observation of direct dipolar exchange processes~\cite{dePaz-Chrom-QM-2013,yan2013:dipole-dipole,Hazzard2014} as well as energy shifts due to long-range interactions in a recent realization of the extended Bose--Hubbard model~\cite{Ferlaino-Extended-2015}.

In all these cases where other relevant energy contributions are scaled down below modest dipolar interaction energies, the overall energies are low and dynamical timescales are long, but the key physical phenomena driven by dipolar interactions remain on display. However, with all other considerations being equal, a larger dipolar interaction makes systems more robust in the face of residual uncontrolled variations of system parameters in space (e.g. harmonic confinement or magnetic field gradients) and in time (noise on magnetic fields, electric fields, or optical potentials), and may be particularly important when they play a role in enabling higher-order processes~\cite{Tewari-EmergenceDipolar,Pair-Hopp-Majorana-Cheng-2011}.

While the several order-of-magnitude differences in $a_{dd}$ and $V_{nn}$ paint a stark contrast between magnetic atoms and polar molecules, some fine details make these differences a bit less extreme. For example, the lanthanoids, which have primary laser-cooling transitions at shorter visible wavelengths approaching the near-UV, can be straightforwardly trapped in optical lattices of smaller spacing, with $\lambda = 532$~nm leading to a factor of 8 enhancement in $V_{nn}$. For molecules, the calculated $V_{nn}$ are based on the maximum electric dipole moment achieved in the limit of full alignment to large applied electric fields. The field-induced electric dipole moments that have been demonstrated thus far have all fallen well short of their maximum electric dipole moments~\cite{ni2010,Takekoshi:STIRAP2014,Molony:STIRAP2014,Park-NaK-GS-2015,NaRbGas}. Furthermore, considering the transition dipole moments between opposite-parity rotational states, which exist even at zero electric field, the corresponding $V_{nn}$ are reduced by a factor of 3 (with a value of $\sim$104~Hz observed for KRb molecules~\cite{Hazzard2014}). Lastly, the conversion of doubly-occupied sites of magnetic atoms into homonuclear Feshbach molecules or ground state molecules allows in principle for an additional factor of 4 enhancement in the achievable magnetically dipolar interaction strengths~\cite{Frisch-Mols-2015}.

It is also important to note that there are many additional dipolar systems under experimental investigation, including nitrogen vacancy centers in diamond~\cite{dolde-nv-2013,Plenio-simulator-2013}, nuclear spins in solid-phase nuclear magnetic resonance (NMR) studies~\cite{Barrett-dipolar,Kaiser-Science-2015}, oscillating charged ions~\cite{Wilson-DipDip-Ions-2014}, and Rydberg atoms interacting via F\"{o}rster resonances~\cite{Pfau-Forster-12,Browaeys-DipDip-14,Browaeys-ExcitationTransfer-15}. These systems offer a range of alternative attributes in the study of dipolar behavior and for quantum simulation studies based on long-range interactions. However, we now move on to a more detailed discussion of magnetic atoms and polar molecules, which are of the most direct relevance to the study of dipolar quantum gases.

\subsection{Magnetically dipolar atoms}

Atoms possess hyperfine structure due to the coupling of their total electronic angular momentum $J$ with a non-zero nuclear spin $I$, resulting in a total angular momentum $\mathbf{F = J + I}$. In a weak magnetic quantization field of strength $B$, the hyperfine energy levels are split up into a manifold of Zeeman substates, the energies of which separate linearly with the applied field $B$. These Zeeman substates are defined by their different magnetic quantum numbers $m_F$, relating to their projection of $\mathbf{F}$ onto the quantization axis, which range from $-F$ to $F$ in unit intervals. These Zeeman sublevels are characterized by different magnetic moments $\mu = m_F g_F \mu_B$ and a linear energy splitting $\Delta E = m_F g_F \mu_B B$ at low fields (with $g_F$ the hyperfine Land\'{e} g-factor~\footnote{Where $g_F = g_J \frac{F(F+1) - I(I+1) + J(J+1)}{2F (F+1)}$, $g_J = 1 + (g_s - 1)\frac{J(J+1) - L(L+1) + S(S+1)}{2J(J+1)}$ is the gyromagnetic ratio due to the electron angular momentum, $g_s \approx 2.0023193$ is the spin g-factor of the electron, and $L$ and $S$ are the atom's electronic orbital and spin angular momentum.}).

\begin{figure}[b]
\includegraphics[height=9cm]{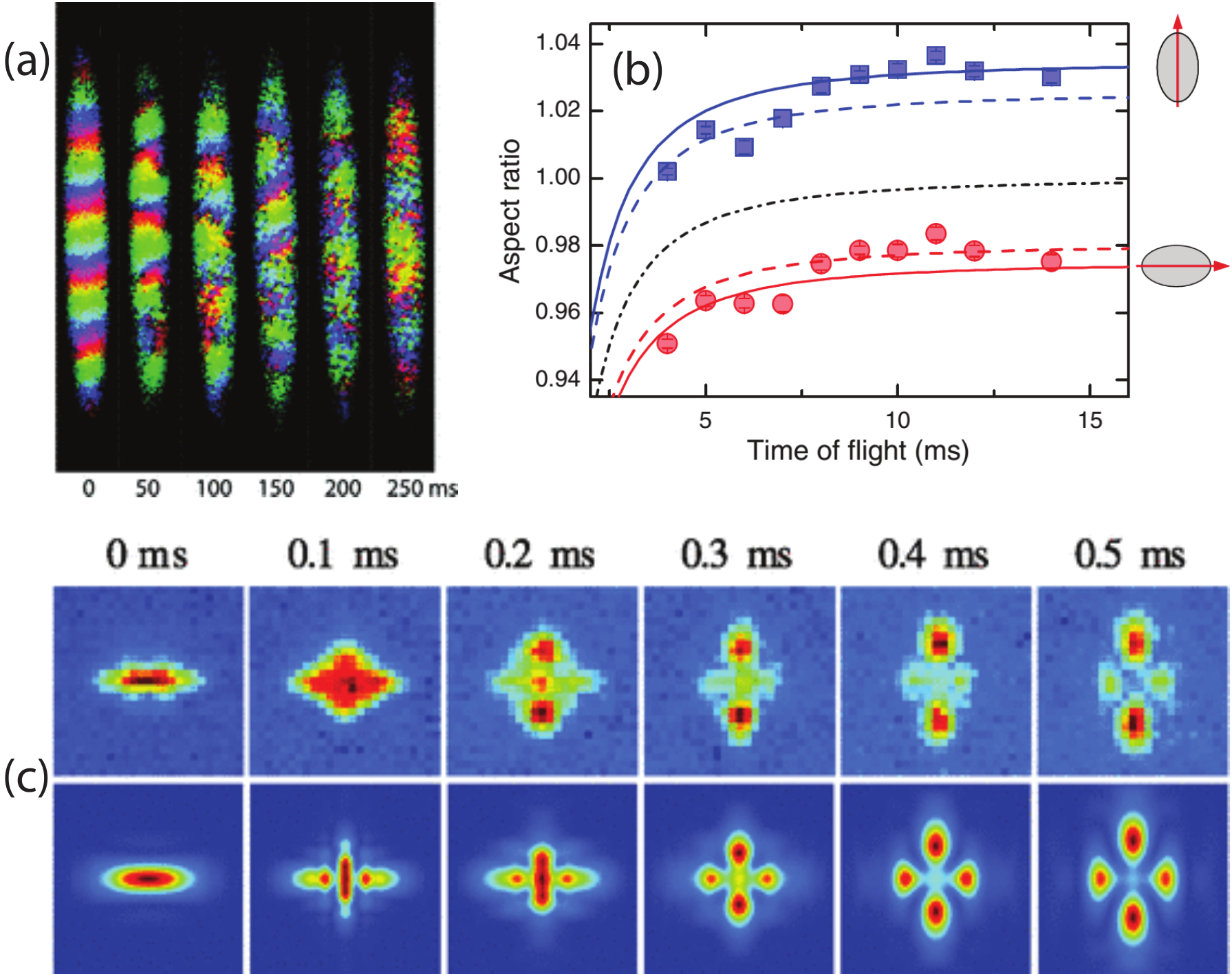}
\caption{\label{Fig:dipolar_effect}
Dipolar effects observed in different ultracold atomic systems. (\textbf{a})~The spontaneously modulated spin textures observed in a dipolar spinor BEC of rubidium. Reprinted figure with permission from~\cite{Vengalattore2008}. Copyright (2008) by the American Physical Society. (\textbf{b})~The deformation of a typically isotropic Fermi surface due to magnetic dipole-dipole interactions in erbium. From~\cite{Aikawa:Er_Fermi_surface2014}. Reprinted with permission from AAAS. (\textbf{c})~The $d$-wave time-of-flight expansion pattern appearing due to collapse of an unstable dipolar BEC of chromium. Reprinted figure with permission from~\cite{Lahaye:Cr_collappse2008}. Copyright (2008) by the American Physical Society.
}
\end{figure}

The magnitude of the magnetic moment $\mu$ varies largely across the different atomic species. For alkaline earth metal atoms such as strontium and calcium, and for alkaline earth-like atoms such as ytterbium, $^{1}S_0$ electron configurations result in zero electronic magnetic moment. For the alkali metal atoms, most widely used in cold atomic physics and first brought to the quantum degenerate regime~\cite{Anderson1995,Davis1995,DeMarco1999}, the presence of a single valence electron leads to magnetic moments of roughly $1 \ \mu_B$. There exist a number of particularly ``magnetic'' atoms with larger magnetic dipole moments. These include chromium ($\mu = 6 \ \mu_B$), which was the first magnetic atom brought to quantum degeneracy in 2005~\cite{ChromiumBEC-2005}, as well as many atoms from the lanthanoid family, including holmium (9 $\mu_B$), erbium (7 $\mu_B$), and dysprosium (10 $\mu_B$).

Magnetic dipole-dipole interactions (MDDIs), which scale as $\mu^2$, are typically negligible in the case of alkali atoms as they are dominated by $s$-wave interactions and other contributions to the total energy. In the $|F,m_F \rangle=|1,1\rangle$ ground state of $^{87}$Rb, for example, the dipole moment of $\mu \approx -\mu_B / 2$ leads to a dipolar length of only $a_{dd} = 0.18 \ a_0$, almost three orders of magnitude smaller than the $s$-wave scattering length. However, dipolar interactions have played a key role in some experiments on spinor Bose--Einstein condensates. Dynamical evolution of the local magnetization or pseudospin (hyperfine state) in a multicomponent atomic quantum gas can be influenced by a number of factors - at the single-particle level by any (pseudo)spin-dependent trapping potentials, and through atomic interactions by spin-dependent and spin-independent $s$-wave interaction strengths, as well as by MDDIs. In the case of mixtures of $^{87}$Rb atoms in different magnetic sublevels, the intraspecies and interspecies scattering lengths all take on a nearly isotropic value of $\sim 100 \ a_0$. Because the contribution to the $s$-wave interaction energy is essentially independent of the (pseudo)spin degrees of freedom, the weak MDDIs can play a significant role in determining evolution of the atomic spin distribution~\cite{Yi2004,Kawaguchi2006,Yi2006}. Described by the general form given in Eq.~\ref{equ:dipdip1} for unpolarized dipoles, MDDIs between atoms have been observed to cause spontaneous formation of short-range spin domains when starting from long-wavelength helical spin textures~\cite{Vengalattore2008,Eto2014}, as shown in Fig.~\ref{Fig:dipolar_effect}~(a).

Dipolar interactions generally play a much larger role for the different magnetic species of atom that have been realized in the quantum regime, such as chromium~\cite{ChromiumBEC-2005,Cr-Fermi}, erbium~\cite{aikawa:bose-einstein_Er_2012,Aikawa-Fermi}, and dysprosium~\cite{lu:strongly_Dy_2011,lu:quantum_Dy_2012}. We briefly highlight some of the important experimental observations that have already been made in these different magnetically dipolar systems, and point out that much exciting dipolar physics is yet to come. In all magnetic quantum gases, the presence of dipolar interactions have been readily observable through anisotropic deformation of expanding clouds of atoms, either through condensate expansion in the case of bosons~\cite{Stuhler:Cr_BEC2005,gries-compare,lu:strongly_Dy_2011} of by deformation of the Fermi surface in the case of fermions~\cite{Aikawa:Er_Fermi_surface2014}, as shown in Fig.~\ref{Fig:dipolar_effect}~(b). The MDDIs have also been seen to modify the collective modes of trapped gases~\cite{Bismut:Cr_collecive2010}, as well as their elementary excitations~\cite{Bismut-ExcSpec-2012}.

The relative strength of dipolar interactions, as compared to $s$-wave interactions, has been enhanced in a number of experiments through the use of magnetic Feshbach resonances~\cite{Inouye-Feshbach,Chin2010}. In early experiments on $^{52}$Cr, this Feshbach control~\cite{Werner-Fesh-Cr} allowed for many important observations, including the first achievement of a strongly dipolar quantum fluid~\cite{lahaye:strong_chromium_2007}, detailed studies of collapse due to dipolar attraction~\cite{Lahaye:Cr_collappse2008,Metz-collapse-2009} (shown in Fig.~\ref{Fig:dipolar_effect}~(c)), as well as stability against dipolar collapse in harmonic traps~\cite{Koch-Stability} and in optical lattices~\cite{LatticeStability-Muller}. More recently, a large number of narrowly-spaced Feshbach resonances have been found for both erbium~\cite{Frisch-Chaos} and dysprosium~\cite{Baumann-Feshbach-Dy,Broad-Fesh-Dys-2015}, and have been employed in $^{164}$Dy to observe the formation of self-organized crystal structures in a dipolar BEC~\cite{Pfau-Crystal-2015}.

Lastly, the confinement of dipolar particles in optical lattices opens up a number of exciting possibilities. For one, lattice confinement can be used as a tool to control the kinetic energy scales of atoms, both coherent (tunneling) and incoherent (thermal). Because dipolar interactions survive in the limit of zero tunneling, while still being of physical consequence in non-polarized samples, this has recently allowed for the observation of nonequilibrium quantum magnetism in a lattice-confined sample of bosonic chromium atoms~\cite{DePaz:Cr_double_well2014}. More recently, the same system has been used to study rich spin dynamics across a range of particle mobilities~\cite{bruno-nonequ-2-2015}. In the case of itinerant spin-polarized samples, long-range interactions can lead to a number of interesting phases such as charge density waves and supersolids~\cite{Sansone-2Dphases-2010,Scarola-SSphase-2006,Buchler-SS}, can lead to interlayer pairing mechanisms analogous to Bardeen-Cooper-Schrieffer (BCS) pairing~\cite{Baranov-DipolePairing,Baranov-Bilayer-image}, and in frustrated geometries can be used to study emergent gauge fields~\cite{Tewari-EmergenceDipolar}. Recently, signatures of nearest-neighbor off-site interactions have been observed with $^{168}$Er atoms in a three-dimensional cubic lattice~\cite{Ferlaino-Extended-2015}, an important first step towards the observation of novel dipolar quantum phases in optical lattices.


\subsection{Electrically dipolar molecules}

Systems of polar molecules are of key experimental interest to a diverse set of scientific applications~\cite{carr:cold_2009}. Before extolling their promise for quantum simulation studies, we briefly discuss a number of other areas of research that would benefit from improved methods for the cooling, trapping, and manipulation of polar molecules.

Tests of fundamental physical theories, such as the Standard Model, are usually performed under high energy conditions at large-scale particle accelerators. However, precision measurements made in the low energy sector may also be sensitive to physics not predicted by the Standard Model. In particular, it has been shown that spectroscopy of polar molecules aligned in modest external electric fields ($\sim 10$~kV/cm) can be used to test and constrain theories that go beyond the Standard Model~\cite{SANDARS1965194,Meyer-molEDM}. This amazing fact results from the extremely large internal electric fields ($\sim 100$~GV/cm) that can be generated within a field-aligned molecule, allowing molecules to act as an extreme laboratory environment for their bound electrons. Such conditions allow for spectroscopic signatures of a possible non-vanishing electric dipole moment (EDM) of the electron, which would imply charge parity (CP) violation at a level inconsistent with the Standard Model of particle physics. The search for new physics requires extremely precise measurements, and meaningful constraints are already being placed on extensions to the Standard Model~\cite{ACME:EDM2014}. Improved methods for the cooling and trapping of polar molecules will allow for longer interrogation times, more precise spectroscopic measurements, and even tighter constraints on a possible value of the electron EDM~\cite{Hudson:EDM2011,ACME:EDM2014}.

Besides precision measurement, cold polar molecules also offer an extremely unique system with which to study coherent and state-controlled quantum chemistry~\cite{silke2010}. Polar molecules can be prepared in particular internal quantum states of molecular rotation and vibration, as well as of their nuclear spin. Their center of mass motion can be cooled to energies well below the corresponding energy spacing between different rovibrational states, and to a regime where only one or a few partial waves contribute to binary molecular collisions. The detailed study of state-controlled reactions, and in particular of the product states of ultracold chemical reactions, would offer unique insight into the challenging dynamical processes involved in chemical reactions.

From the standpoint of quantum simulation and the study of dipolar physics, polar molecules offer many interesting features, including anisotropic interactions, scalable system sizes, a large and rich manifold of internal states, and the possibility of coherent lattice tunneling~\cite{Wall-Mol-Hubbard}. The primary motivating factors, however, are that molecules support very large and tunable electric dipole moments, for molecular states that are inherently stable to single-particle decay (as compared to, e.g., Rydberg atoms). As shown in Table~\ref{tab:title}, many species of molecule offer nearest-neighbor interaction energies $V_{nn}$ in the few to few tens of kHz range, dominating over the typical range of tunneling energies in optical lattice experiments, and approaching or exceeding typical lattice band-gap energies. These extremely strong dipolar interactions may prove especially important when trying to realize higher-order correlated tunneling events enabled by dipolar interactions~\cite{Tewari-EmergenceDipolar}. Still, much progress needs to be made in the experimental study of polar molecules - particularly in the cooling of different species to the quantum gas regime - to harness their considerable potential. Here, we provide a bit of basic information on the physics of polar molecules, and then move on to a discussion of new techniques being applied to the cooling and trapping of molecules.

\begin{figure}[t]
\includegraphics[height=9cm]{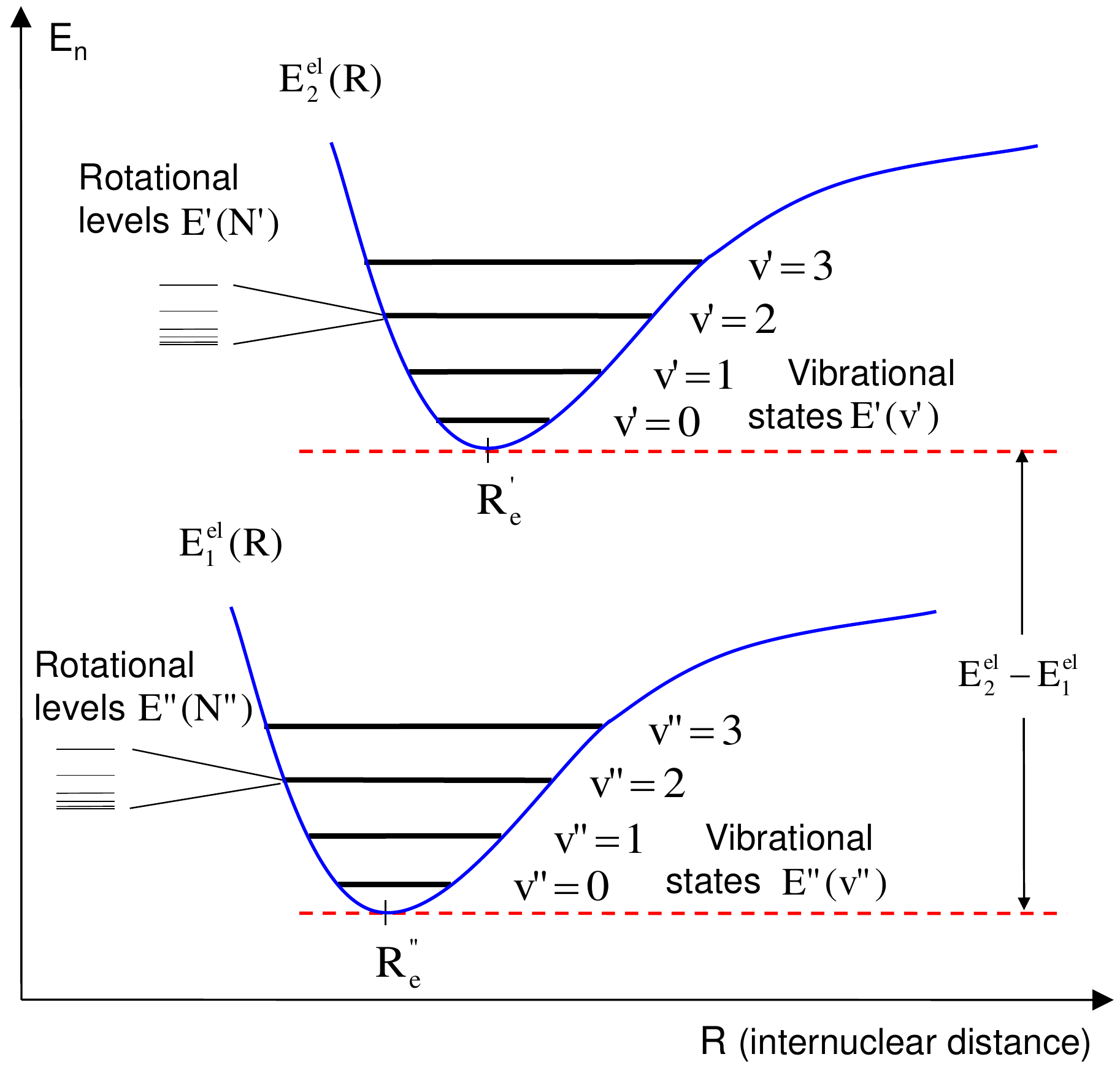}
\caption{\label{Fig:molenergies}
Diagram displaying the characteristic energy levels $E_n$ of different electronic, vibrational, and rotational states of a molecule, shown as a function of internuclear distance $R$.
}
\end{figure}

In general, molecules offer a much richer and more complex energy level structure as compared to atoms, with many additional internal degrees of freedom. As shown in Fig.~\ref{Fig:molenergies}, the level structure of molecules may roughly be divided, in order of descending energy, into electronic, vibrational, rotational, and nuclear (hyperfine) degrees of freedom~\cite{molsBasic}. The transition energy for electronic excitations can be estimated in an extremely coarse fashion by comparing to the hydrogen ground state ionization energy,
\begin{equation}
E_{el}\sim e^2/4 \pi \epsilon_0 (2a_0) \approx 13.6 \ \mathrm{eV} \ .
\end{equation}
Typical electronic excitations have energies $E_{el}$ of a few eV (near UV, optical, and near IR frequencies on the order of a few hundred THz), deviating from this crude approximation due to variations in nuclear charge, electron screening and quantum defects, and different principal quantum numbers and orbital quantum numbers involved in the excitation.

The next largest energy scale in the internal molecular level structure results from vibrations in the relative positions of the atomic nuclei. For the low-lying vibrational levels of closed-shell (and some open-shell) diatomic molecules, the molecular vibrations can be simply modeled as two masses coupled by a spring force $F = - k \Delta x$, corresponding to harmonic oscillator energies for the $\nu^{\mathrm{th}}$ vibrational excitation
\begin{equation}
E_\nu = (\nu+1/2) \hbar \sqrt{k/M} ,
\end{equation}
where $k$ is the effective spring constant of the molecular bond and $M = M_1 M_2 / (M_1 + M_2)$ is the reduced mass of the nuclei. As shown in Fig.~\ref{Fig:molenergies}, the molecular potential is approximately harmonic for small $\nu$, becoming more anharmonic with more closely spaced energies for larger vibrational excitations. This holds up until some final bound state energy, with higher energies corresponding to free scattering states of the individual atoms. Typical molecular vibrations have energy scales of roughly $E_{vib} \sim \sqrt{m_e / M} E_{el} \sim (10^{-3}-10^{-2}) E_{el}$~\cite{carr:cold_2009}, with infrared transition frequencies ranging from a few to $\sim 150$~THz.

The nuclei may also rotate about the molecule's center of mass, resulting in a spectrum of rotational excitations. For polar molecules with no internal angular momentum, relevant to the heteronuclear bialkali molecules such as $^{40}$K$^{87}$Rb with spin singlet ground states~\cite{Ospelkaus-Hyperfine}, the rotational spectra can be simply modeled as rotations of a rigid rotor. The rotational energies are then given by
\begin{equation}
E_{rot}=\hbar^2\frac{N(N+1)}{2I_m} \equiv B N(N+1),
\end{equation}
where $B$ is the rotational constant of the molecule, $I_m = M R_e^2$ is its moment of inertia, $R_e$ is the bond length, and $N$ is the rotational quantum number (with projection $m_N \in \{-N,-N+1, \ldots , N-1, N \})$ along the quantization axis). For a typical range of bond lengths (order few $a_0$) and reduced masses (order few atomic mass units u), the rotational energies will be on the order $E_{rot}= (10^{-6} - 10^{-2})N(N+1) \ \mathrm{eV}$, featuring transition energies ranging from the microwave to the far infrared.

Finally, molecules can additionally possess a large number of states relating to their total hyperfine structure~\cite{MolHypStruct,MolHyp-Hutson}, with some interactions, such as nuclear spin rotation and direct (and indirect) interaction between the nuclear spins, being unique to molecular species. Again, the hyperfine structure is simplified in the case of spin singlet molecules with zero electronic contribution to the total angular momentum. The hyperfine structure can be almost completely described in terms of the nuclear magnetic moments $I$ of the constituent atoms, with their respective magnetic quantum numbers $m_I$ (projections onto the quantization axis) largely serving as good quantum numbers at sufficiently large magnetic fields~\cite{Ospelkaus-Hyperfine}. While this picture holds to a good approximation, small off-diagonal couplings between molecular rotation and the electric quadrupole-moments of the nuclei can be utilized for the coherent manipulation of molecular hyperfine states~\cite{Ospelkaus-Hyperfine}. Additionally, a weak coupling between the nuclear spin and the rotation of a molecule can break the degeneracy normally found in the rigid rotor spectrum for states $|N,m_N\rangle$ and $|N,-m_N\rangle$, important for experimentally addressing individual rotational state transitions with spectral selectivity~\cite{yan2013:dipole-dipole,Hazzard2014}.

This rich internal landscape of individual molecules, featuring many states with a diverse set of properties, allows experimentalists to choose states appropriate for a desired application. For example, molecular states that are easily polarized and are relatively insensitive to magnetic fields, such as the $^3\Delta$ state of ThO and ThF$^+$, have been identified as particularly good candidates for electron EDM measurements~\cite{Meyer:EDM2008,Leanhardt:EDM2011}. Similarly, spectroscopy of vibrational excitations in simple diatomic homonuclear molecules have been identified as an ideal testbed for measuring temporal variations of fundamental constants, such as the electron-to-proton mass ratio~\cite{Zelevinsky2008:mass_ratio,DeMille2008:mass_ratio}. There has been interest in the use of polar molecules as part of a quantum computing architecture~\cite{Demille-QC-MOL}, and particular internal states with large dipole moments may be selected to enhance molecule-molecule interactions, while other states that interact weakly with other molecules and external fields may be used to preserve coherence. Some molecules even have the interesting property that their electric dipole moment may be tuned widely by choice of vibrational level~\cite{LiSr}.

The achievement of large dipolar interactions is also central to the use of molecules for quantum simulation studies. The largest dipole moments are achieved for molecules with smaller internuclear distances, which result in larger charge asymmetries. This has motivated researchers to create molecules in the vibrational (and also the rotational) ground state, as heteronuclear Feshbach molecules formed through magnetoassociation possess negligible dipolar character. Low-lying rotational states of molecules were identified early on as good candidate pseudospins in the simulation of quantum magnetic systems~\cite{barnett:quantum_2006,gorshkov}.  These rotational excited states experience spontaneous decay on timescales that are orders of magnitude longer than most experiments, a necessary requirement to serve as pseudospin states or qubits. These states are experimentally quite favorable because state manipulation can be performed via dipole-allowed transitions at GHz frequencies. More importantly, large transition dipole moments exist between opposite parity rotational states of molecules even in zero electric field, allowing for the experimental studies of dipolar spin exchange with polar molecules~\cite{yan2013:dipole-dipole,Hazzard2014} described herein. In addition to the simulation of quantum magnets, the study of rotational excitations in dense molecular gases may be used to address many other fundamental questions, e.g. related to quantum impurities~\cite{KBW-BEC,Krems-Polaron,Lemeshko-Angulons} and disorder physics~\cite{Yao2014:localization,Krems-Anderson}.

\subsubsection{Interactions with external fields}

A great deal of control may be exerted over molecules through the application of external electromagnetic fields, both static (dc) and dynamic (ac). As with atoms, the hyperfine structure of molecules allows for their manipulation by external magnetic fields. For molecules with an appreciable magnetic moment, such as those not in a spin singlet, this can be used for magnetic trapping in inhomogeneous magnetic fields~\cite{Sawyer-Magneto-Electric}. For experiments involving multiple internal states, relative state energies may be manipulated through external fields to achieve level degeneracies necessary for the study of dipolar interactions that do not conserve internal angular momentum~\cite{Syzranov2014:SOC}.

In contrast to neutral atoms, polar molecules can also be easily manipulated by dc electric fields of moderate strength ($< 100$~kV/cm). We'll consider rigid rotor-like molecules in their vibrational ground state, relevant to experiments involving ground state heteronuclear bialkali molecules. Polar molecules in general possess a large permanent electric dipole moment $\vec{d}_{\mathrm{int}} = d \hat{n}$ that is fixed in the molecule's reference frame, with magnitude $d \sim 1 \ D$. However, such a dipole moment has no preferred direction in the absence of an applied electric field, and the lab-frame dipole moment is effectively zero ($\langle \vec{d} \rangle = 0$). Thus, for samples of molecules in zero applied field, there is no effect of dipole-dipole interactions to first order if all of the molecules are in the same rotational state. This follows as a consequence of the fact that, for zero electric field, the integral of the dipole operator is zero with respect to states of the same parity. However, transition dipole moments between opposite parity rotational states obeying the dipole selection rule $|\Delta N|\leq 1$ are non-zero even in zero electric field. Thus, dipolar interactions can be studied even at zero electric field using non-polarized samples of molecules prepared in multiple rotational states.

\begin{figure}[t]
\includegraphics[height=7cm]{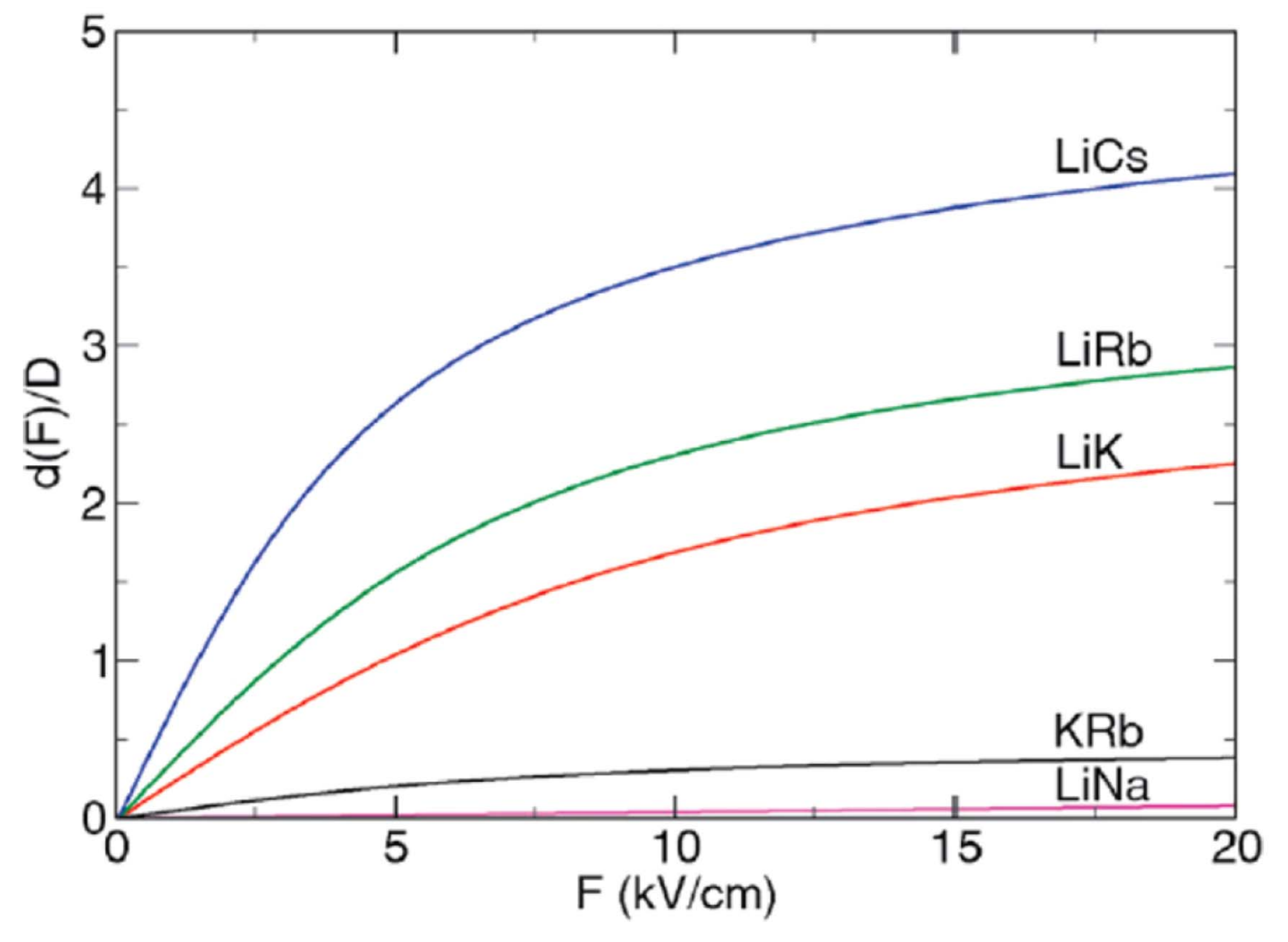}
\caption{\label{Fig:KRB_dipole}
The lab frame electric dipole moment as a function of the applied electric field $F$ for ground state fermionic bialkali molecules. Reprinted with permission from~\cite{Quemener2012:control}. Copyright (2012) American Chemical Society.
}
\end{figure}

For polarized samples, however, large and tunable lab-frame dipole moments can be achieved by application of a dc electric field, which we will set as $\vec{E} = E \hat{z}$ without loss of generality. The interaction of the molecule's dipole with the electric field is described by
\begin{equation}
H_{E}=-\vec{d}\cdot\vec{E} ,
\end{equation}
and leads to alignment of the molecule along the applied electric field. This interaction occurs at second order in the field strength $E$ through off-diagonal mixing of rotational states with opposite parity (with mixing restricted to states obeying $\Delta N = \pm 1$ due to dipole selection rules). The lab frame expectation value of the dipole moment changes linearly with weak electric fields, weak with respect to a critical field given by the ratio of the molecule's rotational constant and permanent dipole moment $E_c \equiv B/d$, such that the Stark energy shifts defined by $H_E$ scale as $\Delta E \propto E^2$. For fields larger than $E_c$, the magnitude of the lab frame dipole moment saturates to the maximum value $d$ as the molecule approaches full field alignment. Figure~\ref{Fig:KRB_dipole} shows the variation of the lab frame dipole moment with applied electric field for various fermionic bialkali molecules, showing a sizeable variation in accessible values of $d$~\cite{Quemener2012:control}.

Similar to the case of neutral atoms, time-varying electromagnetic fields associated with external optical (laser), microwave, and radiofrequency fields allow for a large amount of added control over molecules. Laser fields with spatially inhomogeneous intensity profiles may be used for the creation of conservative optical traps and optical lattices for cold molecules through off-resonant ac Stark shifts (as well as for driving internal state transitions or for absorption imaging~\cite{Wang-Imaging}), with a few key differences as compared to ground state neutral atoms. For one, the presence of vibrational excitations at infrared frequencies offers control in a frequency range not typically accessible in atoms. For simple rovibrational ground state molecules in zero electric field, the ac Stark shifts due to fields near-resonant with electronic transitions are similar to the case of ground state neutral atoms. More generally, however, the polarizability $\alpha(\omega)$ of a polar molecule is a rank-two tensor, described by tensor, vector, and scalar interactions with the external field~\cite{brian}. Considering low-lying rotational states $|N,m_N\rangle$, for example, molecules have an anisotropic polarizability~\cite{brian} that depends on both the polarization of the electromagnetic field and the electronic wavefunction associated with the rotational state. We discuss in more detail the experimental consequences of this feature in Sec.~\ref{sec:sec3exp}. In short, it allows for the creation of internal state-dependent potentials by a different mechanism as compared to the case of neutral atoms, but also leads in general to an unwanted state-dependence that can limit the coherence of pseudospin states relevant to the study of quantum magnetism~\cite{brian,yan2013:dipole-dipole,Hazzard2014}.

In addition to the trapping of molecules with optical potentials, there has also been much interest in the use of intense microwave fields for the engineering of deep, large volume traps~\cite{DemilleUwaveTrap,Hinds-uWaveTrap,Kajita}. Moreover, the ``dressing'' of internal states through near-resonant or far-detuned microwave fields offers another interesting tool by which to manipulate polar molecules. Similar to the case of neutral atoms~\cite{Gerbier-Dressing,Raman-Dressing}, far-detuned polarized microwave fields may simply be used to shift the energy levels of different rotational states, allowing access to level degeneracies relevant to the observation of spin-orbital coupling~\cite{Ueda-EDH,Syzranov2014:SOC}. It has also been suggested that mixing in character from different internal states can directly modify the nature of the dipole-dipole interactions~\cite{Buchler-Crystal-2007}, such as to control the shape of the interaction potential or to allow for the study of novel interactions and topologically non-trivial systems~\cite{gorshkov,Gorshkov2011:tjmodel,Yao-FlatBand-2012,salvatore,Yao2013:Chern}. While detailed experimental studies have been performed on the field alignment and ac polarizability of ultracold molecules, the full power of microwave dressing techniques are as of yet unharnessed.

\subsection{Producing cold polar molecules}

For all of the described applications, it is desirable to be able to produce cold samples of molecules. For precision measurement, a cold source of molecules will reduce Doppler frequency shifts and will enhance interrogation times. For quantum chemistry, the ability to control the thermal kinetic energy, or more generally the collisional kinetic energy~\cite{Walraven-scattering}, when studying molecular collisions allows for unique prospects in the study of reaction dynamics and molecular energy landscapes. Lastly, the quantum simulation of many-body physics with polar molecules is largely predicated on the ability to produce molecular samples at ultracold (nK to $\mu$K) temperatures and at high phase space densities. As in the case of atoms, both bosonic~\cite{Anderson1995,Davis1995} and fermionic~\cite{DeMarco1999}, the achievement of quantum degeneracy will be an important milestone in the study of polar molecules~\footnote{The direct synthesis of molecules from atoms in an optical lattice~\cite{Moses-LowEntropy} is already able to achieve entropies (filling fractions) that would correspond to quantum degeneracy in an equilibrium harmonically trapped gas.}. Driven by these exciting prospects for studying new physics with ultracold molecules, as well as many as of yet unknown applications, there has been much excitement and rapid development in the cooling and trapping of molecules over the past few decades. We now describe two complementary paths that have been explored for the achievement of cold molecular gases: direct cooling and indirect methods. These two strategies seek to either directly cool samples of molecules or to create molecules from samples of cold atoms, respectively.

\subsubsection{Direct cooling of molecules}

Direct cooling methods start from a collection of molecules, and use electric fields, magnetic fields, optical laser fields and additional tools to perform molecule slowing and cooling. Some of the most commonly used techniques for slowing and cooling neutral molecules, as well as the state-of-the-art capabilities of these methods, are listed in Table~\ref{tab:cool}. Just as for many atomic species, conservative methods to slow a collection of molecules, reducing its overall kinetic energy but not increasing its phase space density, must first be performed to enable additional cooling and trapping techniques. There exist multiple techniques for slowing samples of molecules, several of which are listed in Table~\ref{tab:cool}. One such method is Stark deceleration~\cite{Bethlem:stark1999,Bethlem:stark2000,Bochinski:Stark2003,StarkReview,Momose-Stark}. Compared to atoms, molecules feature large electric dipole moments, and can experience large Stark energy shifts in relatively low dc electric fields. Stark deceleration was one of the earliest molecule slowing techniques developed~\cite{Bethlem:stark1999,Bethlem:stark2000}, and cooling can also be achieved through supersonic expansion. With typical molecular dipole moments of a few D, and for \emph{E} fields of order 100~kV/cm, the forward (longitudinal) kinetic temperature of a molecular sample (pulse) can be slowed by tens of K in a single deceleration stage, and molecules can be essentially stopped with a few tens of slowing stages. The technique of Stark deceleration allows for slowing to a few mK, suitable for some methods of trapping. This has in recent years allowed for the evaporative cooling of slowed and trapped $^*$OH radicals~\cite{Stuhl:evaporation2012} to low temperatures and relatively high phase space densities. A related technique for the slowing of molecules is based upon magnetic deceleration~\cite{Hogan:zeeman2007,Narevicius:Zeeman2008,Softley-Zeeman}. Here, inhomogeneous and pulsed magnetic fields of order 1~T are used to slow molecules that have sizable magnetic moments ($\sim 1 \ \mu_B$), allowing for slowing to and trapping at temperatures at around 100~mK with a multistage magnetic ``coilgun'' slower. Optical dipole forces may also be applied to the slowing of fast molecules. In particular, pulsed lasers have been used to create extremely deep optical potentials, used for the single-stage reduction of translational energy by up to 15\%~\cite{Fulton:optical2006}.

\begin{table}[b]
\centering
  \begin{tabular}{|M{7.0cm}|M{2.5cm}|M{2.5cm}|M{2.5cm}|}
\hline
Method & Temperature & Number & PSD \\
 \hline
Buffer gas cooling~\cite{Weinstein:buffer1998,David:neon2009,Lu:buffer2011,Hutzler:buffer_review2012,Hemmerling-Buffer} & $\sim0.4$ -- $4.2$ K & --- & ---\\
Magnetic deceleration~\cite{Hogan:zeeman2007,Narevicius:Zeeman2008,Softley-Zeeman} & $\sim$100 mK & --- & ---\\
Stark deceleration~\cite{Bethlem:stark1999,Bethlem:stark2000,Bochinski:Stark2003,StarkReview,Momose-Stark} & $\sim$1 mK -- 1 K & --- & ---\\
\hline
Stark $+$ evaporation~\cite{Stuhl:evaporation2012} & 5.1 mK & $\lesssim 10^6 \ ^*$ & $\gtrsim 3 \times 10^{-7} \ ^*$\\
Sisyphus cooling~\cite{Zeppenfeld:sisyphus2012,Rempe-uK} & 420 $\mu$K & $3 \times 10^5$ & ---\\
Laser cooling / MOT~\cite{Shuman:SrF2010,Barry:SrF2012,Hummon:YO_2DMOT2013,Barry:SrF_MOT2014,Zhelyazkova:CaF2014,McCarron:2015,Doyle-CaF-MOT,switchingMOT} & 400 $\mu$K & $2 \times 10^3$ & $1.5 \times 10^{-14}$\\
\hline
\end{tabular}
\caption{Direct methods for slowing and cooling molecules. A summary of some of the most commonly used methods for slowing and cooling samples of neutral molecules. For each method, when available and applicable, the typical or state-of-the-art values obtained for temperature, molecule number, and phase space density are listed.
\newline\newline
$^{*}$Indirectly determined
}
\label{tab:cool}
\end{table}

One powerful technique that can in principle be applied to a large variety of atom and molecule is buffer gas cooling~\cite{Weinstein:buffer1998,BufferPrimer,Hutzler:buffer_review2012,Hemmerling-Buffer}. Here, a large reservoir of noble gas atoms confined to a cell and cryogenically cooled to low temperatures (typically 4~K for He or 14~K for Ne~\cite{David:neon2009}) is used to sympathetically cool, through elastically thermalizing collisions, a sample of molecules. In the diffusive limit, with a buffer gas temperature $T_{bg}$, molecules of mass $m_{mol}$ can be cooled to thermal velocities
\begin{equation}
v_{eff}=\sqrt{2k_B T_{bg}/m_{mol}}.
\end{equation}
With one stage 4~K cell, molecules have been cooled to velocities of 140~m/s~\cite{Hutzler:buffer_review2012}. Further cooling can be achieved by adding another cooling cell stage, achieving velocities of 70~m/s~\cite{Lu:buffer2011}. In addition to cooling the translational motion of molecules, their internal motion - vibration and rotation of the nuclei - can also thermalize with the buffer gas. For molecules, vibrational level spacing is typically of order 100~K, with rotational states spaced by only a few K. Thus, in cooling from room temperature down to 4~K, molecules can be well prepared in a single vibrational level and in a few low-lying rotational states. The low temperatures and high flux rates that can be achieved with buffer gas-cooled molecules make them a favorable starting point for additional cooling methods. The technique of buffer gas cooling is especially exciting because it can be applied to an extremely wide variety of atoms and molecules~\cite{BufferPrimer,Hutzler:buffer_review2012,Hemmerling-Buffer}.

While the described methods based on spatially or temporally inhomogeneous electric, magnetic, and optical fields provide a method to remove translational kinetic energy from molecular samples, they in general do not remove entropy or directly increase the molecular phase space density. Given the large success in direct cooling of many atomic species by laser addressing of strong cycling transitions, it would be natural to seek similar capabilities for the cooling of molecules. However, the rich internal structure of molecules complicates the situation severely. Upon driving an electronic transition, a molecule can decay into a number of different vibrational and rotational states from whence it started, such that a closed cycling transition is in general absent. Still, some well-chosen molecules possess ``quasi-closed'' transitions, where their Franck-Condon factors are almost perfectly diagonal (no off-diagonal coupling between different states), such that they have a very high return probability upon electronic excitation~\cite{Rosa:lasr_cooling2004}. By further selecting a rotational level transition of $N=1$ to $N=0$, decay to the original $N=1$ level is the only parity-allowed path~\cite{Stuhl:laser_cooling2008}. Very good Franck-Condon factors make the decay rates of the excited $\nu'=0$ state to vibrationally-excited levels ($\nu\geq3$) of the electronic ground state extremely small ($<10^{-5}$). Thus, with only 2 or 3 repumping lasers, the molecules can scatter $10^5$ photons before decaying to dark states, enough to achieve laser-cooling. So far, SrF~\cite{Shuman:SrF2010,Barry:SrF2012}, YO~\cite{Hummon:YO2013, Yeo:2015, Collopy:2015}, and CaF~\cite{Zhelyazkova:CaF2014,Doyle-CaF-MOT} molecules have all been experimentally laser-cooled, and many more candidates are under consideration, including MgF, BaF, RaF, YbF, TiF, CaH, SrH, TlO, SrBr, and SrCl. Moreover, great recent progress has been made in achieving a magneto-optical trap (MOT) of molecules~\cite{Barry:SrF_MOT2014,McCarron:2015,Doyle-CaF-MOT}, based mainly on a traditional (static) MOT scheme that happened to feature a weak spring constant of trapping due to complex energy level structure~\cite{Tarbutt:2015}. More recently, a time-varying MOT, or ``switching'' MOT, has been demonstrated to achieve molecule temperatures as low as $400\ \mu K$ and densities as high as $6\times 10^4\ \mathrm{cm}^{-3}$~\cite{switchingMOT}. The continued advances and recent breakthroughs in the laser cooling of molecules make this a very promising path to attaining ultracold temperatures. Future extensions to more general classes of molecules may even be enabled new techniques based on cooling with ultrafast laser pulses~\cite{Kielp-ultrafast-mol}. A recent proposal~\cite{Jayich:2014} suggesting the use of frequency-chirped laser pulses to perform cooling by adiabatic rapid passage promises to allow for enhanced diagonal decay probabilities, loosening restrictions on the types of molecules that can be laser-cooled.

One other way to address the lack of closed electronic transitions in molecules, which restricts the number of photons that may be scattered before population is lost to dark states, is to remove more energy and entropy per photon and cool much more efficiently. Several variations of this general methodology exist. One possible method is based on strong bichromatic optical forces~\cite{bichro-hal-1993}, involving two laser frequencies for the slowing and cooling of molecules~\cite{Bichro-Ed-2011}. A single-photon method based on informational cooling has been applied to atoms and suggested for the cooling of molecules~\cite{Price-singlephoton}. Recently, cycled Sisyphus cooling of molecules~\cite{Zepp-PRA,Comparat:sisyphus2014} has been shown to be an extremely powerful method. Roughly speaking, Sisyphus cooling proceeds by having molecules move within an internal state-dependent potential with cycled transitions between the two states (e.g., \emph{A} and \emph{B}). For example, a particle in internal state \emph{A} loses lots of kinetic energy as it moves from its trapping minimum up a large potential ``hill''. Near the highest energy region of this potential, population is transferred from internal state \emph{A} to internal state \emph{B}, for which the potential is much more shallow. The particle then gains less kinetic energy than was originally lost as it moves back down the hill to the trap minimum. At this point, repumping of population back to state \emph{A} will restart the cycle, allowing for a continued decrease in kinetic energy. Molecules lose energy during each cycle, and highly efficient cooling can be made possible if the potentials for states \emph{A} and \emph{B} are quite different. Experimentally, optoelectrical Sisyphus cooling has been explored for CH$_3$F molecules~\cite{Zeppenfeld:sisyphus2012} loaded from a novel centrifuge decelerator source~\cite{Cherenkov-decel}, and cooling from 390~mK to 29~mK has been demonstrated, leading to an increase in phase space density (PSD) by a factor of 29. More recently, this technique has been refined and applied to formaldehyde molecules (H$_2$CO), achieving temperatures as low as 420(90)~$\mu$K and up to four orders of magnitude increase in PSD~\cite{Rempe-uK}.

Lastly, once molecules are slow enough and cold enough to be conservatively trapped, further methods for cooling may be applied. In some cases, evaporative cooling of molecules can be utilized much like for neutral atoms, allowing for enhanced phase space densities through energy-selective particle removal and rethermalization~\cite{EvapField1,EvapField2,EvapField3}. Efficient rethermalization and cooling may also be enabled by the large dipole-dipole interactions between polarized polar molecules~\cite{Quemener:collision2013,Zhu:evaporation2013}, similar to the recent demonstration in dipolar atoms~\cite{Aikawa-Fermi}. Already, evaporative cooling has been applied to samples of Stark-decelerated and trapped $^*$OH radicals~\cite{Stuhl:evaporation2012}. The state-of-the-art capabilities of direct cooling by magneto-optical trapping and laser cooling, optoelectrical Sisyphus cooling, and evaporative cooling are summarized in Table~\ref{tab:cool}.

\subsubsection{Indirect methods for producing cold and ultracold molecules}

Currently available methods for the direct cooling of molecules can typically only achieve mK temperatures (with temperatures as low as $\sim$400$\ \mu$K recently reported~\cite{switchingMOT,Rempe-uK}), and are still many orders of magnitude away from achieving phase space densities in the quantum regime. Presently, only indirect methods for the production of ultracold molecules, based on creating molecules from quantum degenerate atomic gases, have been able to reach the ultracold regime. This strategy relies on the simple energy level structures and scattering properties of neutral atoms, allowing for efficient cooling of atomic samples to high phase space density. It then seeks to transform dilute equilibrium gases of atoms, having typical particle densities $\sim 10^{11}-10^{13} \ \mathrm{cm}^{-3}$ and interparticle spacings $\sim 0.5-2 \ \mu$m, into dilute gases of rovibrational ground state molecules with sub-nm bond lengths. Bridging the gap between free atoms and bound molecules clearly presents several challenges. Foremost, there is a very clear wavefunction mismatch between the tightly bound molecules and the typical free atom scattering state, relating to roughly 4 orders of magnitude difference in internuclear separation. Second, as motivated in Fig.~\ref{Fig:molenergies}, molecules possess a very large internal state space. Efficiently transforming two free atoms into a single state of rotation and vibration (as well as electronic and nuclear degrees of freedom), maintaining a low internal state entropy, thus poses yet another challenge. Third, to preserve ultralow kinetic energies, the very large internal binding energy of the ground state molecules (of order $k_B \times 10^4$~K for KRb~\cite{JinYephystoday}) needs to be removed in an extremely precise fashion. Many additional complications exist, including sensitivity to the atom-atom, atom-molecule, and molecule-molecule scattering properties. As we now describe, over the past decade researchers have developed powerful techniques to overcome these daunting challenges.

\begin{figure}
\includegraphics[height=7cm]{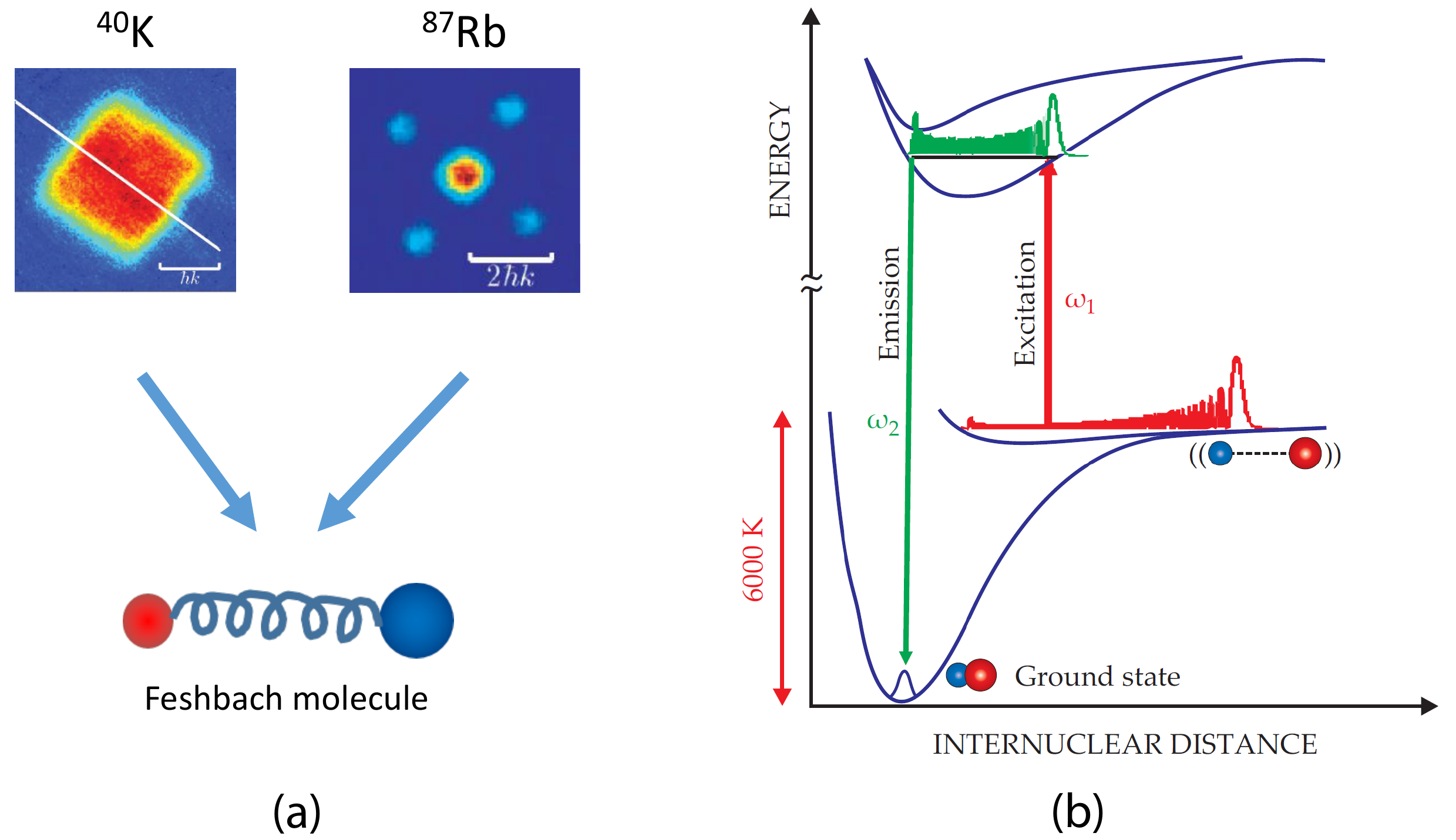}
\caption{\label{Fig:ultracold}
Indirect production of ultracold molecules from ultracold atoms. (\textbf{a})~A mixture of $^{87}$Rb and $^{40}$K atoms is prepared in an optical lattice, from Ref.~\cite{Moses-LowEntropy}. The $^{40}$K atoms form a band insulator (momentum-space image of band-mapped atoms), while the $^{87}$Rb atoms form a Mott insulator (image shows coherent momentum-space interference peaks when the gas is superfluid). By ramping the magnetic field across a Feshbach resonance, $^{40}$K and $^{87}$Rb atoms on a single lattice site are converted into a loosely bound Feshbach molecule. (\textbf{b})~Two-photon stimulated rapid adiabatic passage is used to coherently transfer the Feshbach molecules to the molecular rovibrational ground state. Reproduced with permission from~\cite{JinYephystoday}. Copyright (2011), American Institute of Physics.
}
\end{figure}

The earliest experiments on the production of molecules from cold and ultracold atoms employed the technique of photoassociation (PA)~\cite{Miller-PA-1993,Wynar11022000,Jones:PA_review2006,RbK-Gould}. In PA, two free atoms are excited by laser light in a dipole-allowed transition to a bound state of the electronic excited state potential. By careful choice of the excited state, being short-lived and with a large decay probability to ground state manifolds, deeply bound and even rovibronic ground state molecules can be created through PA~\cite{Fioretti:PA1998,Haimberger:PA2004,Viteau:PA2008}. The probability of excited state transfer is largely influenced by the spatial wavefunction overlap of the free atom scattering state with the excited state (Franck--Condon factor). The typical internuclear separations for free atoms are quite large ($\sim 10^4$~ground state bond lengths) as compared to the vibrational bound states, however this size mismatch may be ameliorated by first creating loosely bound molecules through magnetoassociation~\cite{Pellegrini:FeshbachPA2008,Tolra:FeshbachPA2003}. Here, free atom scattering states may be adiabatically connected to loosely bound Feshbach molecules by sweeping the magnetic bias field across a Feshbach resonance, or rather by radiofrequency or microwave association into bound molecular states near a Feshbach resonance~\cite{Kohler:Feshbachreview2006,Chin2010}. Figure~\ref{Fig:ultracold}~(a) depicts the formation of KRb Feshbach molecules from an ultracold sample of potassium and rubidium atoms (taken from Ref.~\cite{Moses-LowEntropy}).

PA has found much success in the formation of ground state molecules of many species, even allowing the study of cold collisions of molecules~\cite{Hudson-RbCs-collisions}. However, undesirable effects, including irreversible losses of population from the excited state, have so far kept it from producing molecular gases near the quantum degenerate regime. Fortunately, alternative schemes~\cite{STIRAPmolecules,Juha-STIRAP,HeinzenSTIRAP} that largely avoid direct population of electronic excited states, based on two-photon rapid adiabatic passage (STIRAP) were developed and used to coherently convert Feshbach molecules into rovibrational ground state molecules~\cite{Lang:STIRAP2008,Danzl:STIRAP2008,Ospelkaus:STIRAP2008,ni2008}. As depicted in Fig.~\ref{Fig:ultracold}~(b), STIRAP is able to coherently bridge the large energy gap between Feshbach and ground state molecules by exchanging light from two coherent optical fields with very different photon energies. By using copropagating STIRAP lasers very little kinetic energy is imparted to the molecules, and effectively no momentum is imparted if the atoms are first tightly trapped in an optical lattice. Using the STIRAP method, ultracold samples of polar KRb molecules have been prepared in the quantum regime, both in optical dipole traps~\cite{ni2008} and in optical lattices~\cite{amodsen2012,Moses-LowEntropy}, enabling us to study the rich dipolar interactions of ultracold molecules~\cite{yan2013:dipole-dipole,Hazzard2014}. Recently, the production of rovibrational ground state molecules has also been achieved for the species RbCs~\cite{Takekoshi:STIRAP2014,Molony:STIRAP2014}, NaK~\cite{Park-NaK-GS-2015}, and NaRb~\cite{NaRbGas}. The next several years are likely to see an every increasing diversity of dense ground state polar molecule gases as experiments on heteronuclear atomic mixtures continue to progress~\cite{KCs-Nagerl,RbSr-Schreck-Exp,RbSr-Dulieu-Theory,DipoleMoments-AlkAlkEarth,RbYb-Gorlitz-Exp,RbYb-Theory-Meyer,LiCs,LiNa,LiYb-Gupta-Exp,LiYb-Takah-Exp,LiK-Dieckmann,LiK-Chevy,LiRb-Madison,LiRb-Chen-Exp} Integral to the future production of new varieties of ultracold molecular gases is the continued theoretical guidance~\cite{Dulieu1,Hut1,Hut2,Hut3,Hut4,Julienne-Faraday,Quemener2012:control,Mayle-stickycollision,Cote1,Cote2,Cote3,Cote4,Koto1,Koto2,Koto3,Takekoshi-Towards}
with regards to interspecies scattering, dipolar properties of molecules, molecular structure, and molecular production pathways.

\section{Observing long-range dipole-dipole interactions with polar molecules}
\label{sec:sec3exp}

Here, we discuss the recent observation of nonequilibrium quantum magnetism in a system of lattice-trapped ultracold KRb molecules~\cite{yan2013:dipole-dipole,Hazzard2014}. These experiments illustrate the strength of dipolar quantum gases - despite the use of a very dilute, high entropy sample of molecules unstable to exothermic chemical reactions, innate long-range spin-spin interactions allowed for the observation of coherent dipolar excitation exchange. In the following we describe the techniques that were employed to overcome some of the challenges in this system, as well as the experimental evidences for long-ranged spin-spin interactions driven by dipolar excitation exchange.

\subsubsection{Dealing with chemical loss}

One of the primary challenges in using ultracold KRb molecules for quantum simulation is the fact that two colliding (rovibrational) ground state KRb molecules are unstable to inelastic loss, most likely through the exothermic chemical reaction $\mathrm{KRb} + \mathrm{KRb} \rightarrow \mathrm{K}_2 + \mathrm{Rb}_2$~\cite{silke2010}. Moreover, KRb molecules are also unstable to two-body collisions with K atoms that remain in the system, through the process $\mathrm{K} + \mathrm{KRb} \rightarrow \mathrm{K}_2 + \mathrm{Rb}$~\cite{silke2010}. The presence of such inelastic loss pathways may just be an unfortunate detail particular to KRb and other molecules with exothermic chemical reaction pathways, irrelevant to ultracold polar molecules that enjoy endothermic chemical reactions~\cite{Takekoshi:STIRAP2014,Molony:STIRAP2014,Park-NaK-GS-2015,NaRbGas}. However, KRb molecules also experience shortened lifetimes in the presence of Rb atoms, with the corresponding process $\mathrm{Rb} + \mathrm{KRb} \rightarrow \mathrm{Rb}_2 + \mathrm{K}$ being endothermic~\cite{silke2010}. It is believed that even two chemically stable particles (where, e.g., at least one is a molecule with complex structure) can form long-lived collision complexes that greatly enhance the probability for inelastic three-body recombination~\cite{Mayle-stickycollision}, and some evidence exists for these ``sticky collisions'' in samples of (endothermic) RbCs molecules~\cite{Takekoshi:STIRAP2014}. Thus, whether in the form of direct two-body loss or enhanced three-body loss, inelastic collisions are likely to be relevant to future experiments involving ultracold molecules. Here we describe some aspects of chemical loss in ultracold KRb, and provide details of how this loss was mitigated in order to observe coherent dipolar excitation exchange.

Incoherent particle loss is typically considered to be a detriment to quantum simulation studies, and investigations are generally restricted to cases where the particle lifetimes greatly exceed the timescales of physical interest. Under ultrahigh vacuum conditions and for typical particle densities ($n \sim 10^{11}-10^{13} \ \mathrm{cm}^{-3}$), lifetimes on the order of tens to hundreds of seconds are regularly achieved, limited mainly by single particle loss due to background gas collisions and by inelastic three-body collisions~\cite{3bodylossdalibard}. For distinguishable (by internal state) KRb molecules, it was experimentally found~\cite{silke2010} that barrierless chemical reactions have a two-body decay coefficient of $\beta_{\mathrm{exp}} = 1.9(4) \times 10^{-10} \ \mathrm{cm}^3/\mathrm{s}$, leading to extremely short sample lifetimes ($\ll 1 \mathrm{s}$) for typical experimental densities. This loss rate was close to the predicted universal value $\beta_{\mathrm{uni}} = 2 (h/M)\bar{a} = 0.8 \times 10^{-10} \ \mathrm{cm}^3/\mathrm{s}$ (with $M$ the reduced mass of the colliding molecules and $\bar{a}$ the characteristic length scale of the Van der Waals potential relating to the molecule-molecule collision~\cite{Julienne-Faraday}).

\begin{figure}
\includegraphics[height=7cm]{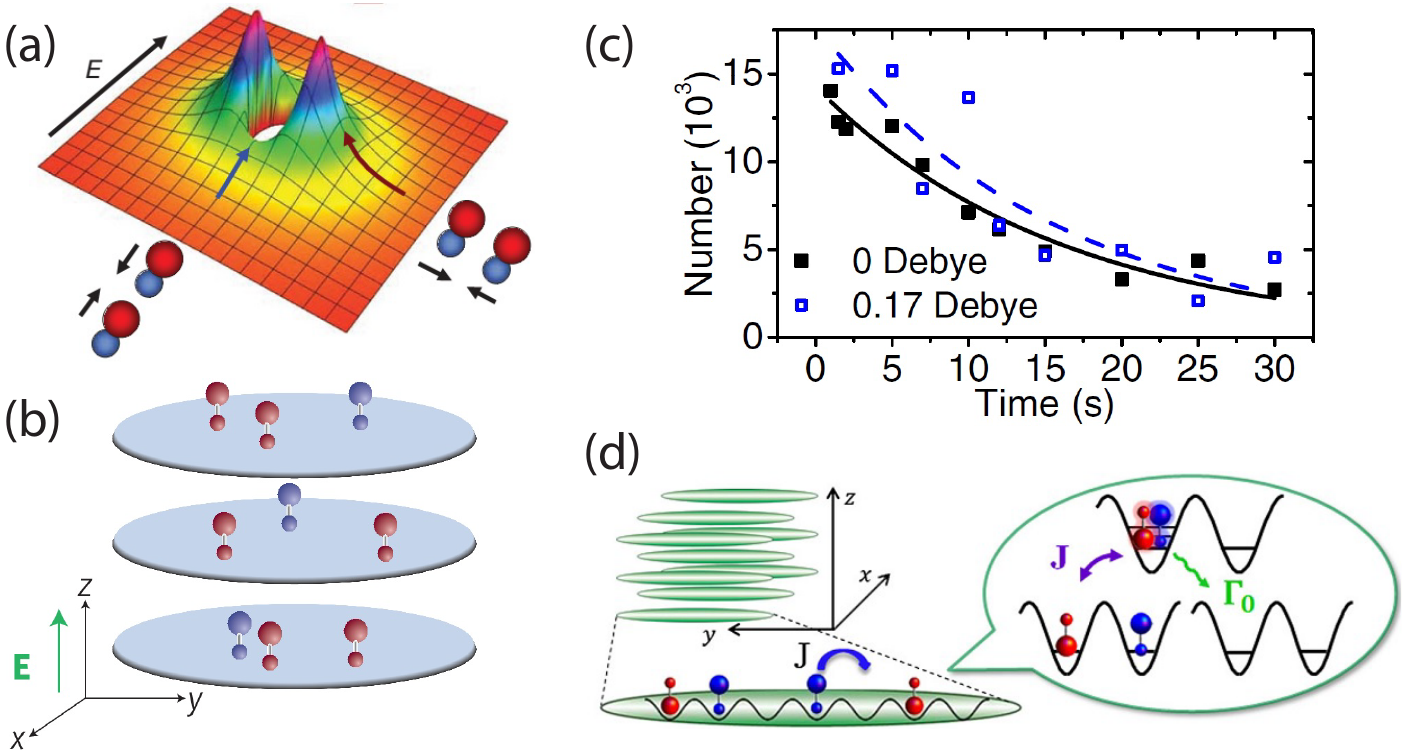}
\caption{\label{Fig:Losses}
Methods for suppressing chemical loss of KRb molecules. (\textbf{a})~For collisions between identical dipolar fermionic molecules in an applied electric field $E$, the $p$-wave energy barrier suppressing chemical loss is raised for repulsive ``side-to-side'' collisions and is lowered for attractive ``head-to-tail'' collisions. Reprinted by permission from Macmillan Publishers Ltd: Nature~\cite{ni2010}, copyright~(2010). (\textbf{b})~By confining polar molecules to a two-dimensional geometry, with dipoles oriented out of the plane of motion, collisions between identical fermionic molecules can be largely stabilized. Adapted by permission from Macmillan Publishers Ltd: Nature Physics~\cite{miranda2011}, copyright~(2011). (\textbf{c})~Long lifetimes, limited only by off-resonant light scattering, are achieved when molecules are pinned in a deep three-dimensional optical lattice~\cite{chotia:long-lived_2012}. When tunneling is negligible, the sample lifetime is independent of the molecular details (shown for molecules at zero electric field and for molecules with a moderate dipole moment in an applied field). (\textbf{d})~Suppression of losses for mobile molecules due to the continuous quantum Zeno effect~\cite{zhu2014:quantum_zeno}. When the on-site loss rate $\Gamma_0$ greatly exceeds the tunneling rate between neighboring sites $J$, tunneling-induced loss is strongly suppressed.
}
\end{figure}

While the lifetimes for samples of distinguishable molecules are extremely short, the quantum statistics of the fermionic $^{40}$K$^{87}$Rb molecules actually lead to a large suppression of the two-body loss coefficient in spin-polarized samples. In the collision between identical fermionic molecules, the requirement that the total wavefunction describing the collision be antisymmetric leads to the absence of low-energy $s$-wave collisions. For identical KRb molecules, the lowest energy symmetry-allowed $p$-wave collision channel has a centrifugal energy barrier of $k_B \times 24 \ \mu\mathrm{K}$ . The two-body loss coefficient for identical particles is thus greatly suppressed in samples with thermal energies much lower than this $p$-wave barrier, and a suppression by about two orders of magnitude was observed in dense samples of KRb at temperatures of 250~nK~\cite{silke2010}. The dipolar interactions between molecules can significantly alter the two-body collision process, in effect lowering or raising the collisional energy barrier depending on whether the interaction between the molecules is attractive or repulsive. This effect is shown in Fig.~\ref{Fig:Losses}~(a), with attractive ``head-to-tail'' collisions lowering the energy barrier and repulsive ``side-to-side'' collisions raising the barrier (taken from Ref.~\cite{ni2010}). These modifications to the collision process only take effect when the KRb molecules are imbued with a lab frame dipole moment in an applied electric field. When both ``side-to-side'' and ``head-to-tail'' collisions are possible in a trapped molecular sample, enhanced losses due to ``head-to-tail'' collisions lead to shortened lifetimes in applied electric fields.

In Ref.~\cite{miranda2011}, researchers were able to greatly suppress (by two orders of magnitude) the influence of lossy ``head-to-tail'' collisions, allowing for stable gases of polar KRb molecules with large dipole moments. To achieve this suppression, the molecular motion was confined to only two spatial dimensions by loading into the lowest band of a one-dimensional optical lattice, forming an array of highly oblate pancake-shaped traps as shown in Fig.~\ref{Fig:Losses}~(b). By applying an electric field such that the dipole orientation was out of the plane of these two-dimensional traps, ``head-to-tail'' collisions were geometrically disallowed.

Unfortunately, this enhanced stability due to the suppression of $s$-wave collisions and confinement in reduced dimensions largely does not extend to mixtures of molecules in different internal (e.g., hyperfine or rotational) states, or to chemically reactive bosonic molecules. Generically, one way to mitigate the effects of inelastic two-body collisions is to simply prevent separated particles from ever coming together. This brute-force approach, which can be accomplished by simply loading molecules into very deep 3D optical lattices with negligible tunneling, was shown in Ref.~\cite{chotia:long-lived_2012} to be extremely successful. In particular, as shown in Fig.~\ref{Fig:Losses}~(c), extremely long molecule lifetimes on the order of $15-20$~s can be achieved independent of the molecular details. These achieved lifetimes are simply limited by off-resonant light scattering from the intense lattice laser beams.

This method for suppressing chemical loss through quenched tunneling has been employed in the herein described experiments, allowing for the observation of coherent dipolar exchange~\cite{yan2013:dipole-dipole,Hazzard2014} between fixed quantum (pseudo)spins, with two internal rotational states serving as (pseudo)spin states. It would stand to reason that studies of \emph{itinerant} quantum magnetism, featuring both charge (molecule) and spin (internal state degree of freedom) mobility, would be beyond the scope of analog quantum simulation with ultracold KRb molecules, owing to the exothermic chemistry of KRb and the inability to provide an energetic barrier to collisions between distinguishable internal states. However, in the limit that the local on-site chemical loss rate $\Gamma_0$ greatly exceeds the tunneling bandwidth of itinerant molecules, the counterintuitive quantum Zeno effect~\cite{TheRealQuantumZeno} provides a mechanism for stability induced by loss. Roughly speaking, the local dissipation due to on-site chemical reactions and loss acts in a similar fashion to a projective measurement~\cite{Itano1990:quantum_zeno,Streed2006:quantum_zeno}, destroying the coherent evolution of the system by coupling the molecules to a large number of internal and motional states ($\sim k_B \times 15 \ \mathrm{K}$ of energy are released in the reaction). The coherent tunneling is suppressed due to the strong local dissipation, as depicted in Fig.~\ref{Fig:Losses}~(d), and for two neighboring sites the rate of tunneling-induced loss is suppressed as $\Gamma_{\mathrm{eff}} \approx 2 J^2 / \Gamma_0$ in the limit of strong dissipation ($\Gamma_0 \gg J$, with $J$ the inter-site tunneling rate). This continuous version of the quantum Zeno effect has previously been observed for homonuclear Feshbach molecules~\cite{Syassen2008:quantum_zeno}, and we have recently observed the same effect with ultracold mixtures of ground state KRb molecules~\cite{yan2013:dipole-dipole,zhu2014:quantum_zeno}. The on-site loss rates are so large for KRb molecules that the process naturally couples to multiple motional bands of the optical lattice in spite of the large bandgap energy~\cite{zhu2014:quantum_zeno}. Such large loss rates $\Gamma_0$ should ensure the stability of mobile molecules for hundreds of coherent tunneling events (at tunneling rates comparable to the dipolar spin-exchange frequency). There is thus hope that quantum phenomena associated with itinerant magnetic systems may be studied with mobile chemically reactive molecules, stabilized through dissipation~\cite{ZenoPwave}.

\subsection{Experimental system: making molecules behave as quantum magnets}

The aforementioned pinning of molecules in a deep optical lattice allows one to realize a stable system of ultracold KRb molecules~\cite{chotia:long-lived_2012}. Moreover, as we now motivate, it also allows for coherent dipolar spin exchange dynamics to be studied in an extremely high entropy sample of KRb molecules. As previously discussed, the ground state KRb molecules are formed from rubidium and potassium atoms loaded into a deep optical lattice via magnetoassociation followed by STIRAP~\cite{Lang:STIRAP2008,Danzl:STIRAP2008,Ospelkaus:STIRAP2008}. In the experiments we shall describe, the filling fraction of the resultant molecules in the lattice is low ($< 10\%$), with presumably an approximately homogeneous random filling of the lattice sites. For systems of ultracold alkali (pseudo)spin mixtures, in which second-order tunneling processes drive the superexchange interactions between neighboring spin states~\cite{trotzky:time-resolved_2008}, quantum magnetism can typically only be studied at extremely low temperatures and entropies. For entropies similar to those of use in experiments with molecules, there would be no signatures of quantum magnetism. For polar molecules, spin-spin interactions are driven directly by dipole-dipole interactions without the requirement of particle tunneling. Thus, the observation of coherent dipolar spin exchange does not require low motional entropy of the KRb molecules. Instead, by quenching particle tunneling, the large amount of entropy associated with the external degrees of freedom (site occupations) of the KRb molecules is divorced from the initially small amount of entropy associated with the internal (pseudospin) degrees of freedom. Because the molecules are initially created in only a single internal (vibrational, rotational, and hyperfine) state, the entropy associated with the internal degrees of freedom is essentially zero. While rapid progress is being made in the achievement of molecular samples with extremely high lattice filling and low motional entropy~\cite{Moses-LowEntropy}, the ability to faithfully simulate nonequilibrium quantum magnetism even at high entropies is of extreme practical import.

Starting from a low-filling gas of molecules pinned to lattice sites, spin-polarized in the rovibrational ground state (and a single hyperfine state~\cite{Ospelkaus-Hyperfine}), a coherent mixture of internal states is created by resonant driving with electromagnetic fields. For the purpose of studying spin-spin interactions driven by dipolar exchange, low-lying rotational states act as excellent (pseudo)spin states due to their long lifetimes (effectively infinite as compared to experimental timescales), easy manipulation with microwave fields, and the strong dipole-dipole interactions that they support~\cite{barnett:quantum_2006,gorshkov,gorshkov2,hazzard:far-from-equilibrium_2013,Wall-moleculechapter}. Here, we restrict our discussion to effectively spin-1/2 systems with only two internal states, as studied in the described experiments~\cite{yan2013:dipole-dipole,Hazzard2014} (more complex arrangements that take advantage of multiple internal states and microwave dressing are discussed in Refs.~\cite{gorshkov,manmana:topological_2013,Yao2013:Chern,Yao-FlatBand-2012}). We refer to the initially populated rotational ground state as ``spin down'', $|N,m_N\rangle = |0,0\rangle \equiv |$$\downarrow$$\rangle$. Population can then be coherently transferred to a state in the first excited rotational manifold, $N=1$ with $m_N \in \{-1,0,1\}$, which we will refer to as ``spin up'' ($|$$\uparrow$$\rangle$). All of these states support dipole allowed interactions with $|0,0\rangle$ even at zero electric field. The choice over which $m_N$ state will serve as $|$$\uparrow$$\rangle$ can be made through the polarization of the resonant microwaves, or more practically through the frequency of the applied radiation. In the case of a simple rigid rotor, the $|1,\pm 1\rangle$ states will separate in energy from the $|1,0\rangle$ state in an applied electric field~\footnote{To avoid complication, we refer to states at finite electric field by the notation of the zero-field states to which they are adiabatically connected.}. Even in the absence of an applied electric field, weak coupling between the nuclear spin and the rotation of a molecule breaks the degeneracy normally found in the rigid rotor spectrum for different $m_N$ states, allowing for spectral selectivity over the rotational state transitions~\cite{yan2013:dipole-dipole,Hazzard2014}.

The Hamiltonian describing these molecular systems is given simply by a sum over single-particle terms and two-body dipolar interactions. For quenched tunneling, the single-particle terms have a simple single-site formulation. Site-specific energies of the two rotational states are captured by an effective ``magnetic field'' term $h_i \hat{S}^z_i$ (ignoring spin-independent energy terms that are irrelevant in the absence of tunneling), where $\hat{S}^z_i$ is the $z$ spin-1/2 operator for site $i$ defined as $\hat{S}^z_i = ( |$$\uparrow$$_i\rangle \langle$$ \uparrow$$_i| - |$$\downarrow$$_i\rangle \langle$$\downarrow$$_i|)/2$. Additionally, spin raising and lowering operators for site $i$ are defined as $\hat{S}^+_i = |$$\uparrow$$_i \rangle \langle$$\downarrow$$_i |$ and $\hat{S}^-_i  = |$$\downarrow$$_i \rangle \langle$$\uparrow$$_i |$, respectively, obeying the commutation relations $[\hat{S}^z_i,\hat{S}^\pm_j] = \pm \delta_{ij}\hat{S}^\pm_i$. These raising and lowering operators are related to the $x$ and $y$ spin-operators as $\hat{S}^\pm_i = \hat{S}^x_i \pm i\hat{S}^y_i$. The effective field terms $h_i$ have several contributions - the states' field-free energy difference, influences of electric and magnetic fields, and differential ac Stark shifts from the trapping lasers. In addition to the diagonal elements, the homogeneous microwaves used to couple the two rotational states can introduce a transverse field term of the form $\Omega \hat{S}^{x(y)}_i$.

We now describe the two-body dipole-dipole interactions that enable the study of quantum magnetism. Following from Eq.~\ref{equ:dipdip1}, the dipole-dipole interaction between two molecules $i$ and $j$, having dipole moments $\hat{\mathbf{d}}_i$ and $\hat{\mathbf{d}}_j$ and with relative position vector $\mathbf{r}_i-\mathbf{r}_j\equiv r_{ij} \hat{r}_{ij}$, is described by the interaction Hamiltonian
\begin{equation}
\hat{H}^{\mathrm{int}}_{ij}=\frac{1}{4\pi\epsilon_0}\frac{\hat{\mathbf{d}}_i\cdot \hat{\mathbf{d}}_j - 3(\hat{\mathbf{d}}_i\cdot \hat{r}_{ij})(\hat{\mathbf{d}}_j\cdot \hat{r}_{ij})}{r_{ij}^3} \ .
\label{equ:dipdip11}
\end{equation}
While the total angular momentum of interacting dipoles is conserved, the internal (spin) and external (orbital) angular momentum of the dipoles are coupled in a non-trivial way~\cite{Kawaguchi2006,Syzranov2014:SOC}. Specifically, this interaction can be conveniently reexpressed as a sum of terms that exchange $p$ units of angular momentum projection between the external (orbital) and internal (rotational) degrees of freedom~\cite{gorshkov}, taking the form
\begin{equation}
\hat{H}^{\mathrm{int}}_{ij}=\frac{-\sqrt{6}}{4\pi\epsilon_0 r_{ij}^3}\sum\limits_{p = -2}^2 (-1)^p T^2_{-p}(\mathbf{C}) T^2_p (\hat{\mathbf{d}}_i,\hat{\mathbf{d}}_j) \ .
\label{equ:dipdip12}
\end{equation}
Presently, we consider only the $p=0$ terms. Physically, these are the only relevant terms when the there are no degeneracies (single particle or pairwise) of the rotational state energies, such that processes that change the global distribution of rotational state populations are energy off-resonant and strongly suppressed. Thus, the system of effectively spin-1/2 molecules is closed and its net magnetization is conserved. Here, the orbital term is given by $T^2_0(\mathbf{C}) = - (1-3 \cos^2\theta_{ij})/2$, where $\theta_{ij}$ is the angle that $\hat{r}_{ij}$ makes with the quantization axis $\hat{z}_q$ ($\cos\theta_{ij}=\hat{r}_{ij}\cdot \hat{z}_q$). The term relating to the internal spin is given by
\begin{equation}
T^2_0 (\hat{\mathbf{d}}_i,\hat{\mathbf{d}}_j) = \frac{2}{\sqrt{6}}[\hat{d}_0^i \hat{d}_0^j + \frac{\hat{d}_{+1}^i \hat{d}_{-1}^j + \hat{d}_{-1}^i \hat{d}_{+1}^j}{2}] \ ,
\end{equation}
with $\hat{d}_0$ and $\hat{d}_{\pm 1}$ the dipole operators.

\begin{figure}
\includegraphics[height=7cm]{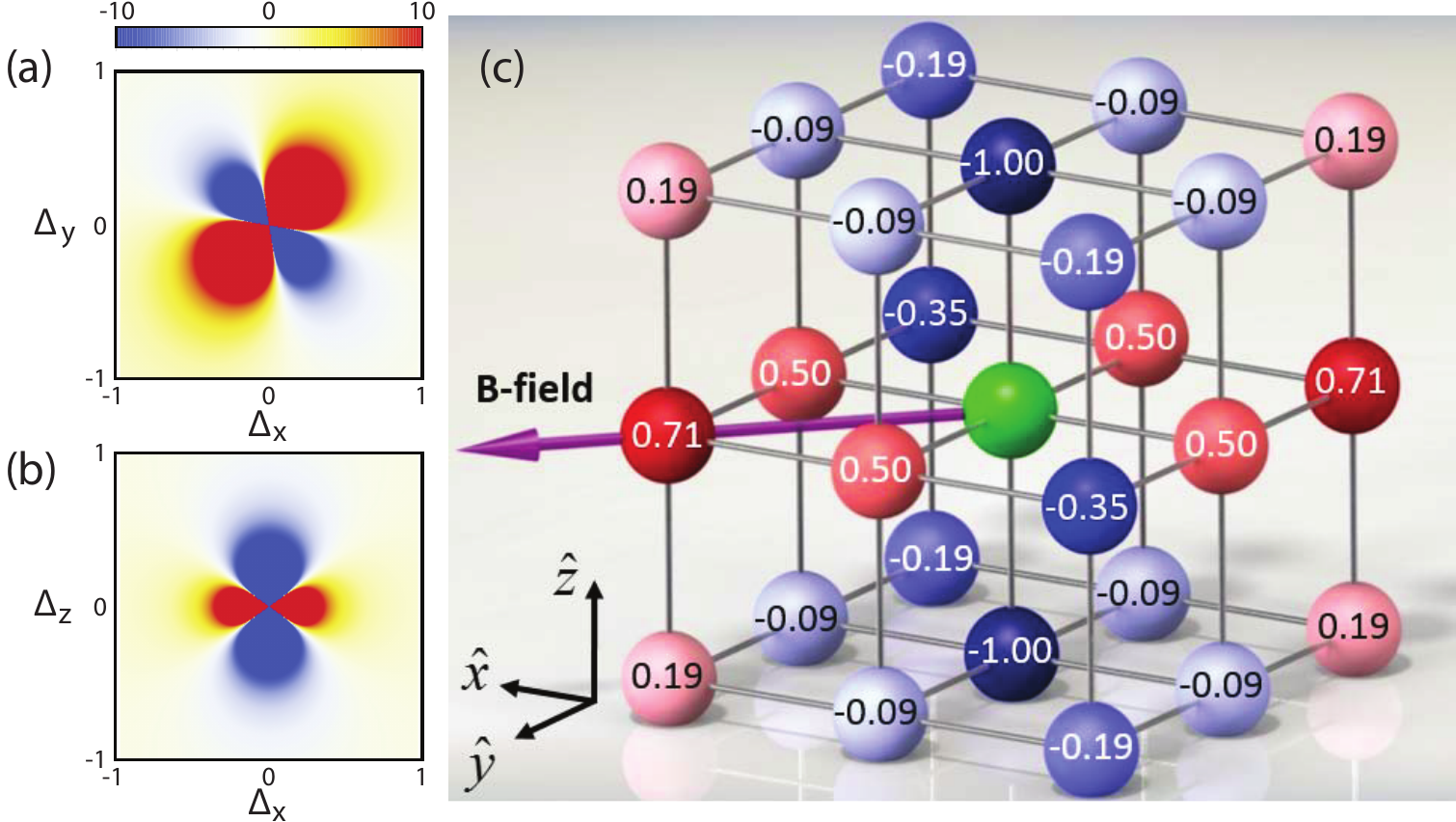}
\caption{\label{Fig:geometric_factor}
The dependence of dipole-dipole interactions on molecular orientation and separation~\cite{yan2013:dipole-dipole}. (\textbf{a},\textbf{b}) Considering a quantization axis along the $(\hat{x}+\hat{y})/\sqrt{2}$ direction as in Refs.~\cite{yan2013:dipole-dipole,Hazzard2014}, we show the dependence of the dipolar geometrical factor $-V^{ij}_{dd}$ on the relative position vector $\mathbf{r}_i - \mathbf{r}_j = \Delta_x \hat{x} + \Delta_y \hat{y} + \Delta_z \hat{z}$ of two molecules $i$ and $j$. In (\textbf{a}), we show for the case $\Delta_z = 0$, and in (\textbf{b}) for the case $\Delta_y = 0$. (\textbf{c}) For discrete positions defined by the sites of a three-dimensional lattice, we illustrate the relative strengths (and signs) of the dipolar geometrical factor $-V^{ij}_{dd}$ felt at several sites surrounding a fixed central molecule (shown in green). Here, all the values are normalized in magnitude to the largest dipolar coupling strength at separations $\Delta_{\{x,y,z\}}=\{0,0,\pm a\}$~\cite{yan2013:dipole-dipole}.
}
\end{figure}

We now explicitly consider how this form of the dipole-dipole interaction can be recast in terms of quantum magnetic interactions, focusing on the case of zero applied electric field explored in Refs.~\cite{yan2013:dipole-dipole,Hazzard2014}. At zero electric field, there is zero dipolar interaction between molecules in the same rotational state (equal parity) due to electric dipole selection rules. However, the transition dipole moment relating to the exchange of rotational excitations between two molecules takes its largest value at zero field. This non-zero off-diagonal coupling between the $|$$\uparrow$$\rangle$ and $|$$\downarrow$$\rangle$ states can be expressed as a direct spin-spin interaction of the form
\begin{equation}
\hat{H}^{\mathrm{int}}_{ij}=V^{ij}_{dd} J_\perp (\hat{S}^+_i \hat{S}^-_j + \hat{S}^-_i \hat{S}^+_j)  \ ,
\label{equ:dipdip13}
\end{equation}
where $V^{ij}_{dd} = (1-3\cos^2\theta_{ij}) / |\mathbf{x}_{i} - \mathbf{x}_j|^3$ is a purely geometrical factor of the dipole-dipole coupling, and $\mathbf{x}_i = \mathbf{r}_i/a$ is the position vector normalized by the lattice spacing $a$. Figure~\ref{Fig:geometric_factor} illustrates the geometrical dependence of the term $V^{ij}_{dd}$, revealing the richly anisotropic and long-ranged nature of the dipolar interactions~\cite{yan2013:dipole-dipole}. The exchange coupling constant between configurations $|$$\uparrow$$_{i}\rangle$$|$$\downarrow$$_{j}\rangle$ and $|$$\downarrow$$_{i}\rangle$$|$$\uparrow$$_{j}\rangle$ is given by
\begin{equation}
J_\perp = \frac{\text{$\langle$$\downarrow$$_i|\langle$$\uparrow$$_j| [ \hat{d}_0^i \hat{d}_0^j + \frac{\hat{d}_{+1}^i \hat{d}_{-1}^j + \hat{d}_{-1}^i \hat{d}_{+1}^j}{2} ] |$$\uparrow$$_i\rangle|$$\downarrow$$_j\rangle$}}{4\pi\epsilon_0 a^3} \ .
\label{equ:exch}
\end{equation}

This exchange coupling constant will be impacted by the choice of rotational excited state. For example, for the choice $|$$\uparrow$$\rangle = |1,0 \rangle$ at zero field, one finds $\langle$$\downarrow$$| \hat{d}_0 |$$\uparrow$$\rangle = \langle$$\uparrow$$| \hat{d}_0 |$$\downarrow$$\rangle = d/\sqrt{3}$, where $d$ is the molecule's permanent dipole moment in the molecular frame. The exchange coupling is then $J_\perp = d^2 / 12\pi\epsilon_0 a^3$, giving an exchange frequency of $\sim 2\pi \times 104$~Hz for these states of KRb molecules ($d = 0.571$~D) in a lattice of spacing $a=532$~nm~\footnote{This estimate of 104~Hz is for the experimental conditions of Refs.~\cite{yan2013:dipole-dipole,Hazzard2014}, where the expected frequency is reduced by a few percent from the simple prediction of $2\pi \times 109$~Hz due to mixing of different nuclear (hyperfine) states.}. This changes if we instead choose a state with non-zero angular momentum projection, e.g. $|$$\uparrow$$\rangle = |1,1 \rangle$ where we have $\langle$$\downarrow$$| \hat{d}_0 |$$\uparrow$$\rangle = \langle$$\uparrow$$| \hat{d}_0 |$$\downarrow$$\rangle = 0$ and $\langle$$\downarrow$$| \hat{d}_{-1} |$$\uparrow$$\rangle = - \langle$$\uparrow$$| \hat{d}_{+1} |$$\downarrow$$\rangle = -d/\sqrt{3}$. Here the sign of $J_\perp$ is reversed, and the magnitude is reduced by a factor of 2 (from the factor of 1/2 in Eq.~\ref{equ:exch}). One can understand this difference by simply thinking of two classical dipoles that rotate together in phase about the quantization axis at the frequency of excitation, experiencing different values of $\theta_{ij}$ at a rate much faster than the scale of interactions. The time-averaged value of their interaction is effectively scaled by the $\theta$-averaged value of $1-3\cos^2\theta$, namely by -1/2. In contrast, this averaging effect is absent for classically oscillating dipoles (analogous to the $p_z$ orbitals of the $|1,0\rangle$ excitation). Clearly, as we discuss in more detail later, this ability to control the magnitude and sign of dipolar interactions through excited state choice can provide a discrete way to tune the strength of interactions. It may also be of future use for more fundamental reasons, for example for the realization of antiferromagnetic exchange coupling in frustrated two-dimensional (2D) lattice geometries.

For the many localized molecules experiencing the described spin-spin interaction, this total combination of single molecule terms ($\hat{H}^0$) and interactions ($\hat{H}^{\mathrm{int}}$) realizes a long-ranged and anisotropic spin-1/2 quantum XY model with a spatially varying longitudinal field
\begin{equation}
\label{eqn:hamiltonianA}
\hat{H}= \hat{H}^0 + \hat{H}^{\mathrm{int}} = \sum_i h_i \hat{S}_i^z + J_\perp\sum_{i < j}\frac{1-3\cos^2\theta_{ij}}{|\mathbf{x}_{i} - \mathbf{x}_j|^3} \left(\hat{S}_i^+\hat{S}_j^- +\hat{S}_i^-\hat{S}_j^+\right) \ .
\end{equation}
In the presence of an electric field (and for homogeneous particle density), additional interactions of the form $J_z \hat{S}_i^z\hat{S}_j^z$ between molecules in the same state allow for realization of the spin-1/2 XXZ model and Heisenberg quantum magnetism~\cite{barnett:quantum_2006,gorshkov,gorshkov2,hazzard:far-from-equilibrium_2013}. Furthermore, allowed particle tunneling provides access to an extremely rich $t$-$J$-$V$-$W$ model of itinerant quantum magnetism with density-density and density-spin interactions~\cite{gorshkov}.

Now that we have motivated how quantum magnetic interactions can in principle be realized with ultracold molecules, we detail in the following sections how this idealized model for quantum magnetism was realized in experiment, and how experimental signatures of long-ranged spin-spin interactions were revealed.

\subsubsection{Dealing with spin-dependent light shifts}

Even absent the spatially varying longitudinal fields of Eq.~\ref{eqn:hamiltonianA}, the equilibrium and dynamical properties of a system described by this Hamiltonian are entirely nontrivial, owing to the long-ranged and anisotropic nature of the dipolar interactions and the three-dimensional arrangement of molecules. Still, the types of phenomena that may be studied are only further enriched through control of the landscape of ``magnetic fields'' $h_i$ that the molecular spins experience. However, as we now discuss, uncontrolled noise or inhomogeneities of the applied electric, magnetic, and laser fields will prove detrimental to our ability to simulate and observe coherent quantum magnetic dynamics.

The most fundamental consequence of static ``field'' inhomogeneity can readily be understood by examining a system of only two isolated spins $i$ and $j$, having local ``fields'' $h_i$ and $h_j$. In general, the energy (field) difference $\delta_{ij} = h_i - h_j$ will act as a spin-dependent site-to-site energy bias, serving to suppress the transport of spin excitations supported by $\hat{H}^{\mathrm{int}}$ at a rate $J_{ij} = J_\perp V^{ij}_{dd}$. In other words, spin-dependent trapping potentials will suppress the range over which spin excitations can propagate. When $\delta_{ij} \gg J_{ij}$, the direct dipolar excitation exchange between molecules $i$ and $j$ is effectively ``shut off'' due to being energy off-resonant beyond the exchange bandwidth. More generally, the exchange dynamics will behave akin to off-resonant Rabi dynamics - the magnitude of exchange will be suppressed, and the exchange frequency will increase as $\tilde{J}_{ij} = \sqrt{J^2_{ij} + \delta^2_{ij}}$. At the very minimum, so that our system supports excitation exchange over even the shortest length scales, the typical nearest-neighbor field difference should be smaller than the nearest-neighbor exchange frequency $\sim J_\perp$. To observe the long-ranged nature of the dipole-dipole interaction, we should furthermore hope, for example, that the typical field difference over two, three, and four sites is less than $\sim J_\perp / 8$, $J_\perp / 27$, and $J_\perp / 64$, and so on. While it is inevitable that finite spatial inhomogeneities of the field terms $h_i$ will cause the $1/r^3$ dipolar interaction to be effectively truncated at some large distance, it would be desirable to ``flatten out'' the effective ``magnetic field'' terms as much as experimentally possible. In these experiments, performed at zero electric field and with a homogeneous offset magnetic field, the main contribution to the ``field'' inhomogeneity is the differential light shifts (ac Stark shifts) of the molecular rotational levels in the intense and inhomogeneous lattice laser fields.

Differential light shifts are a common occurrence in the trapping of neutral atoms. For linearly polarized laser light, differential light shifts between states in different hyperfine manifolds result when the trapping laser frequency is not sufficiently far-detuned (with respect to the states' bare energy difference $\delta E$, usually the hyperfine-splitting of order 1~GHz) from atomic resonance (such as the D$_1$ and D$_2$ lines). This light shift is proportional to $\delta E/\Delta_{light}$, where $\Delta_{light}$ is the detuning of the laser from resonance, e.g. of order 100~THz, such that the differential light shift is very small ($< 0.1\%$), and spatial inhomogeneities of it are even smaller for most traps. Furthermore, by control of laser polarization or by tuning between two strong resonances, a complete cancelation of the differential polarizability can be achieved under ``magic'' conditions. In general, the polarizability of atoms is essentially independent of laser intensity (for experimentally relevant laser intensities), such that perfect cancelation of differential polarizability is achieved at all positions within an inhomogeneous optical trap under ``magic'' conditions.

\begin{figure}
\includegraphics[height=7cm]{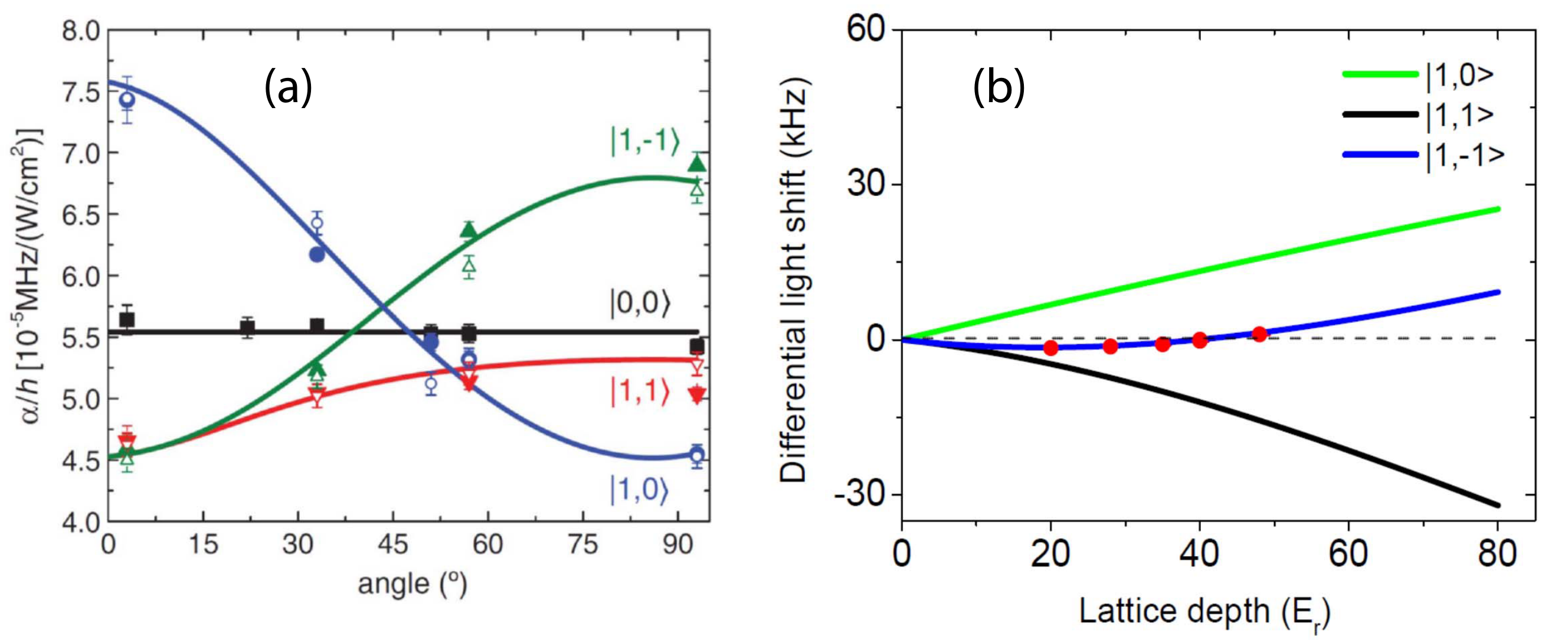}
\caption{\label{fig:polarizability}
(\textbf{a})~The anisotropic polarizability of KRb molecules for different rotational spin states $|N,m_N\rangle$, from Ref.~\cite{brian}. Shown are the ac polarizabilities of the different states trapped in a one-dimensional optical lattice, with respect to the angle that the electric field of the lattice laser light makes with the quantization axis. The states exhibit strong (and different) dependences on the laser polarization, reflecting their anisotropic polarizability and the different electronic wavefunctions of these spin states. Where points coincide for different spin states relates to the existence of a ``magic'' polarization angle, which can minimize the differential light (ac Stark) shift for two states at a given laser intensity. (\textbf{b})~The differential light shifts with respect to the $|0,0\rangle$ ground state as a function of lattice depth (proportional to laser intensity), for molecules trapped in a three-dimensional optical lattice as described in Ref.~\cite{yan2013:dipole-dipole}. Red points show the measured differential light shifts for the $|1,-1\rangle$ spin state, while solid lines are theory plots (no fitting parameters).
}
\end{figure}

This ability, however, is not generally present for molecules. In contrast to most atoms, the dependence of the induced ac polarizability of molecules on the electric field polarization of laser light is highly anisotropic. For the excited ($N \neq 0$) molecular rotational states we consider, this is a reflection of the high anisotropy of their electronic distributions. Such an anisotropy is clearly not present for alkali metal and alkaline earth atoms with outer $s$-shell electrons. Somewhat surprisingly, a lack of polarizability anisotropy is also found for lanthanide atoms with anisotropic inner-shell electronic distributions~\cite{AnisPol-Dulieu}. Figure~\ref{fig:polarizability}~(a), taken from Ref.~\cite{brian}, shows the dependence of the ac polarizability of KRB molecules on the linear laser polarization angle (with respect to a magnetic quantization field axis) for the states $|0,0\rangle$, $|1,-1\rangle$, $|1,0\rangle$, and $|1,1\rangle$, revealing significant anisotropy for the low-lying excited states. A ``magic'' polarization angle can typically be found where two states experience the same ac polarizability - shown for the states $|0,0\rangle$ and $|1,0\rangle$ near $\theta \approx 54^\circ$ ($\cos^2\theta = 1/3$)~\cite{brian}. Unfortunately, however, the polarizability of molecules is also largely intensity-dependent, as shown in Fig.~\ref{fig:polarizability}~(b). Thus, the differential ac Stark shift of two rotational states cannot be simultaneously cancelled at different positions within a spatially inhomogeneous optical trapping potential. While the use of flat-top potentials with homogeneous laser intensity will likely provide a future route to dealing with this issue (and to minimizing spatial variations of differential ac Stark shifts for experiments involving more than two internal states, where a universally ``magic'' polarization cannot be found), the experiments we discuss~\cite{yan2013:dipole-dipole,Hazzard2014} are operated under conditions where the differential light shifts are first-order insensitive to laser intensity, so as to minimize inhomogeneities due to the spatially varying laser intensity in the (Gaussian profile) trapping laser beams.

\subsection{Observing dipolar excitation exchange through microwave spectroscopy}

\subsubsection{Experimental Ramsey spin echo protocol}

In the absence of powerful capabilities for the local preparation, control, and detection of molecular spin (rotational) states, one can still hope to detect the presence of direct dipole-dipole spin exchange processes through global measurements of a large molecular ensemble. In systems of Rydberg atoms, for example, strong interactions between atoms lead to clear spectroscopic shifts and/or broadening of excitations to Rydberg states as compared to the case of single atoms, providing a spectroscopic signature of long-range interactions~\cite{LineBroad1,LineBroad2,RydBlock-Saff,RydBlock-Brow,Schau27032015}. For our KRb molecules, however, analogous shifts or line-broadening will be on the order of the exchange coupling strength $|J_\perp/h| \sim 100$~Hz (for the $|1,0\rangle$ excited state) - small compared to the MHz-level shifts in Rydberg atom systems, but large compared to the local inhomogeneities of the effective ``magnetic field'' terms $h_i$. Still, while the differential light shift can be made small over the range of a few to ten lattice sites, global variations over the $\sim$50-site range (in each direction) of our three-dimensional (3D) molecular ensemble lead to broadening of the microwave transitions on the order of 1~kHz, obfuscating any evidence of dipolar interactions.

An alternative scheme for observing dipolar interactions in systems of cold polar molecules, based instead on Ramsey spectroscopy, was proposed in Ref.~\cite{hazzard:far-from-equilibrium_2013}. In the simplest scenario, every molecule is excited to a superposition of the states $|$$\uparrow$$\rangle$ and $|$$\downarrow$$\rangle$ by a $\pi/2$ microwave pulse, and the spin coherence of the molecules is interrogated after some evolution time by a second microwave pulse. For our large molecular ensembles, the many different interparticle spacings and orientations lead to a large multi-valued spectrum of two-body interaction energies (cf. Fig.~\ref{Fig:geometric_factor}~(c)). As described in Ref.~\cite{hazzard:far-from-equilibrium_2013}, Ramsey spectroscopy should reveal signatures of the dipolar exchange interactions, specifically through an oscillatory decay of the global spin coherence. However, because of the global variation in $h_i$, only a fast decay of coherence on the timescale of $\sim 1$~ms is observed for standard Ramsey spectroscopy. Luckily, global spin echo pulses - $\pi$ pulses that lead to rephasing of the spin evolution in an inhomogeneous (static) field~\cite{Hahn-Echo-1,CarrPurcell-Echo-1} - may be used to mitigate the effects of single-particle dephasing while preserving the signatures of dipolar interactions. Specifically, the single-particle ``magnetic field'' terms ($\hat{H}^0 = \sum_i h_i \hat{S}_i^z$ of Eq.~\ref{eqn:hamiltonianA}) reverse their sign upon inversion of the molecular spins by a $\pi$ pulse, while the pairwise interactions ($\hat{H}^\mathrm{int}$ of Eq.~\ref{eqn:hamiltonianA}) are invariant under such a global spin-flip. This simple Ramsey spin echo protocol, combined with global measurements of the rotational state populations, underlies the main experimental technique used for observing dipole excitation exchange in Refs.~\cite{yan2013:dipole-dipole,Hazzard2014}.

\begin{figure}
\includegraphics[height=8cm]{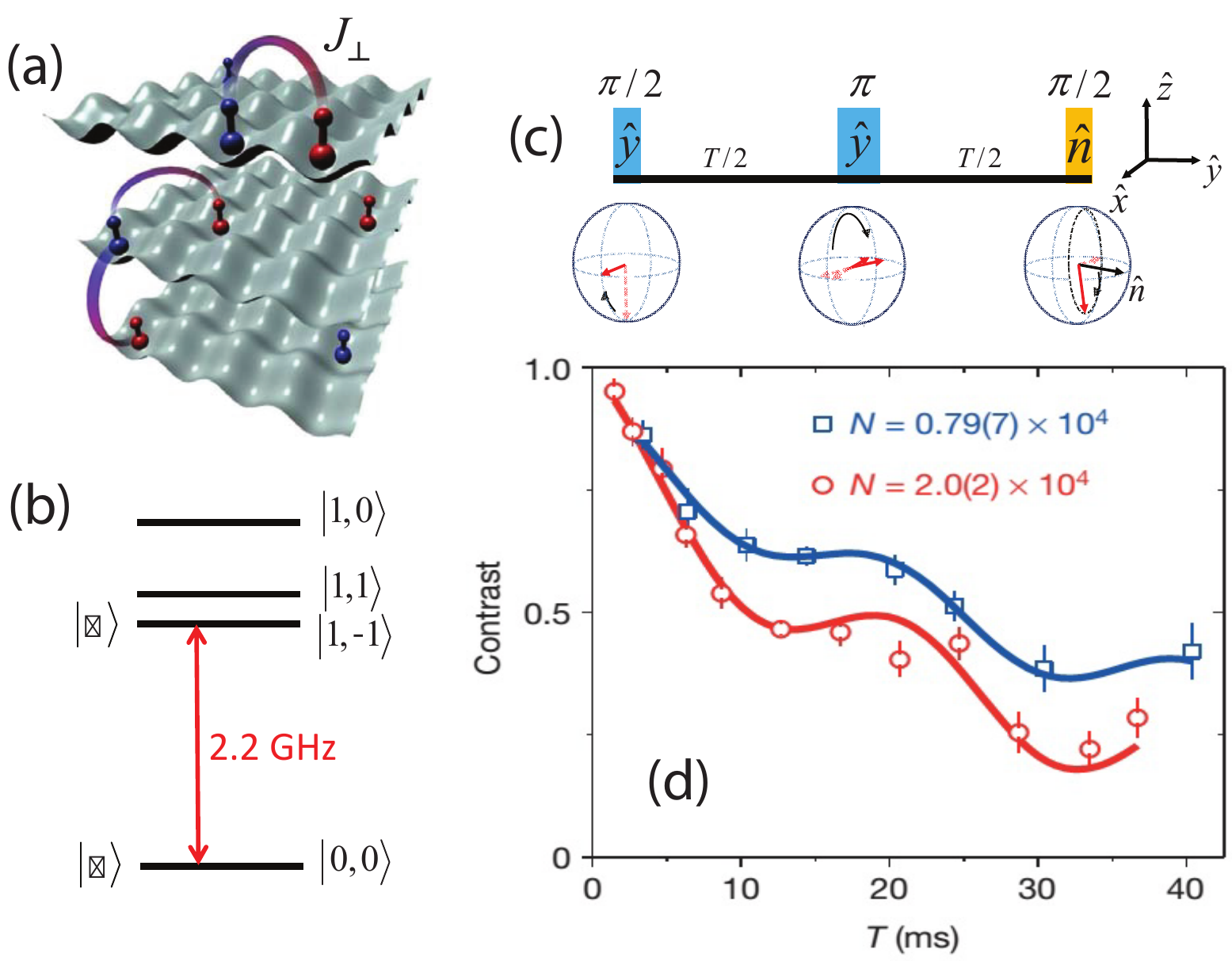}
\caption{\label{fig:scheme}
Dipolar interactions of polar molecules in a 3D lattice (from Ref.~\cite{yan2013:dipole-dipole}).
(\textbf{a})~Polar molecules are formed and trapped in a deep 3D optical lattice, formed by three mutually orthogonal standing waves generated from laser light with wavelength $\lambda = 1064$~nm. Microwaves are used to address the transition between two rotational states (red and blue represent different rotational states). The exchange coupling constant $J_\perp$ characterizes the spin-exchange interaction energy. (\textbf{b})~The schematic energy diagram (not to scale) of the ground and first-excited rotational states.
We use $|0,0\rangle$ and $|1,-1\rangle$ as our two spin states.
(\textbf{c})~Ramsey spin echo pulse sequence used to detect dipole-dipole interactions. A ($\pi/2)_y$ pulse initializes the molecules in a coherent superposition of rotational states.  A spin-echo ($\pi)_y$ pulse sequence is used to correct for effects arising from single-particle inhomogeneities across the sample, mainly from residual light shifts. A final $\pi/2$ is used to probe the molecules' spin coherence.
(\textbf{d})~The Ramsey fringe contrast as a function of interrogation time, shown for several different molecule densities (numbers). The dynamics are characterized by an oscillatory decay, as shown by the empirical fit lines (described in the text).
}
\end{figure}

This microwave Ramsey spectroscopy scheme was used in Refs.~\cite{yan2013:dipole-dipole,Hazzard2014} to study the dipolar interactions of dilute samples of up to $2\times 10^4$ ground-state KRb molecules, prepared in and confined to individual sites of a three-dimensional (3D) optical lattice, as depicted in Fig.~\ref{fig:scheme}~(a). Microwaves with frequency $\sim$2.2~GHz are used to couple the $|N,m_N\rangle = |0,0\rangle$ and $|1,-1\rangle$ rotational states, which form the two-level system of $|$$\downarrow$$\rangle$ and $|$$\uparrow$$\rangle$ states. The degeneracy of the $N=1$ rotational states is broken due to the interaction between the nuclear quadrupole moment and the rotation of the molecules~\cite{hyperfine}. Under the experimental conditions of a $54.59$~mT magnetic bias field (in proximity of a Feshbach resonance used in molecule formation), transitions from $|0,0\rangle$ to the states $|1,0\rangle$ and $|1,1\rangle$ are higher in frequency than the $|1,-1\rangle$ transition by 270~kHz and 70~kHz, respectively, as shown in Fig.~\ref{fig:scheme}~(b). As shown in Fig.~\ref{fig:scheme}~(c), the magnetic quantization field is oriented at 45 degrees with respect to two of the lattice directions (labeled $\hat{x}$ and $\hat{y}$). In this experiment, the lattice laser polarizations are chosen such that spatial variations in the tensor AC polarizabilities of the $|0,0\rangle$ and $|1,-1\rangle$ states across the molecular cloud are minimized~\cite{yan2013:dipole-dipole}.

Starting from a sample spin-polarized in the $|$$\downarrow$$\rangle = |0,0\rangle$ state, coherent microwaves are used to initiate and probe dipolar spin dynamics. Figure~\ref{fig:scheme}~(b) shows a basic spin-echo pulse sequence and its Bloch sphere representation. Starting with the molecules prepared in the $|$$\downarrow$$\rangle$ state, the first $(\pi/2)_y$ pulse (i.e. rotation about the $\hat{y}$ spin axis) creates a superposition state $\frac{1}{\sqrt{2}}(|$$\downarrow$$\rangle+|$$\uparrow$$\rangle)$ at each occupied site. Fidelities of greater than $99\%$ are obtained for $(\pi)_y$ pulses. After a free evolution time $T/2$, we apply a $(\pi)_y$ spin echo pulse, which flips the spins and thus reverses the direction of single-particle precession due to the local fields $h_i$. This echo pulse is used to remove the residual differential AC Stark shift, but has no impact on the dipolar spin-exchange interactions. Ignoring interactions, the spins rephase after another free evolution time $T/2$, at which point we probe the coherence by applying a $\pi/2$ pulse with a phase offset $\phi$ relative to the initial excitation pulse, corresponding to rotation about the spin axis $\hat{n}(\phi) = \cos\phi\hat{x} + \sin\phi\hat{y}$~\cite{martin:Sr_2013}. We then measure the fraction of molecules remaining in the $|$$\downarrow$$\rangle$ state (assuming the total number is approximately constant) as a function of this offset phase, which yields a Ramsey ``fringe'' measurement. The ``visibility'' or ``fringe contrast'' of this Ramsey $\phi$-dependence serves as a signature of the molecules' spin coherence. By varying the total interrogation time $T$, we can study the temporal dynamics of this molecular spin coherence (Ramsey fringe contrast), as shown in Fig.~\ref{fig:scheme}~(d) for several different molecular densities.

The most striking features of the measured contrast curves in Fig.~\ref{fig:scheme}~(d) are the oscillations, along with the overall decay signaling the loss of spin coherence. We attribute both the contrast decay and the oscillations to the presence of dipole-dipole interactions in the molecular ensemble. Simply put, the spectrum of interaction energies derived from the many different interparticle spacings and orientations (set by $V_{dd}$, cf. Fig.~\ref{Fig:geometric_factor}) is broad and multi-valued, being mostly discrete at high energies and densely sampled at low energies. The largest interaction energies, corresponding to nearby molecules, drive the oscillations of the Ramsey fringe contrast at short times. The presence of many different interaction energies, however, leads to the overall decay and absence of revivals in the contrast at long times. In the following, we look in more detail at these two pieces of evidence for dipolar interactions, as well as other direct tests of the nature of the interparticle interactions.

\subsubsection{Density-dependent loss of spin coherence}

The Ramsey fringe contrast dynamics in Fig.~\ref{fig:scheme}~(d), shown for two different molecule numbers relating to different particle densities, hint at a significant density-dependence to the timescales of coherence decay. Earlier, we noted that molecules enjoy an extremely long lifetime when confined to a deep optical lattice, limited to roughly 20~s by inelastic light scattering~\cite{chotia:long-lived_2012}. Although small, the corresponding single-molecule loss rate can exceed the tunneling rate in our deepest optical lattice potentials, and can thus be used to systematically reduce the molecule number while preserving the molecule positions, by simply holding the molecules in the optical lattice prior to performing spin echo Ramsey spectroscopy. Moreover, because the beam waists of our trapping lasers (as well as that of an additional strong laser beam that we turn on to enhance the rate of off-resonant light scattering) far exceed the spatial extent of our molecular cloud, this single-molecule loss is essentially unbiased, removing particles without preference based on their spatial position. Thus, a systematic reduction of the molecule number additionally relates to a systematic reduction of the molecule density (filling factor in the lattice), without introducing significant distortion to the shape of the molecule distribution.

\begin{figure}
\includegraphics[height=7cm]{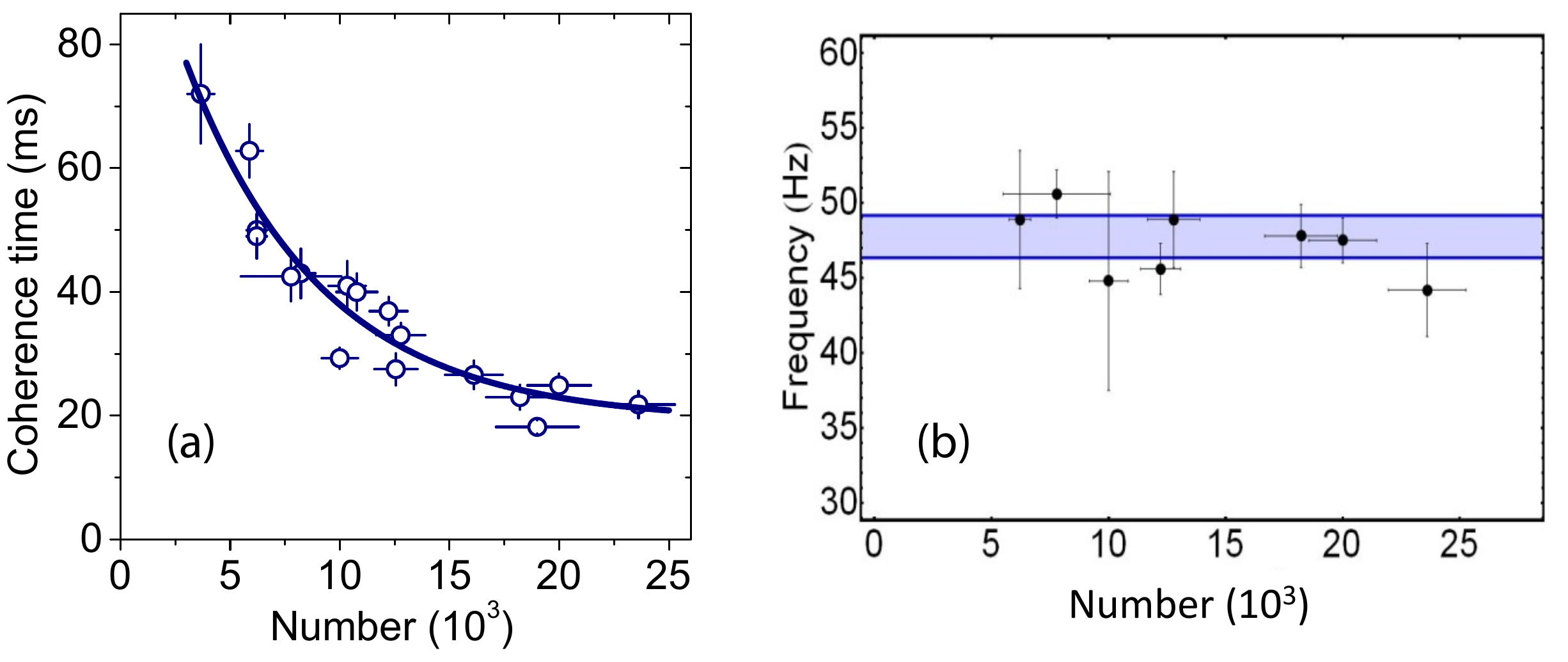}
\caption{\label{fig:density}
Spin coherence dynamics of interacting polar molecules (from Ref.~\cite{yan2013:dipole-dipole}).
(\textbf{a})~The density dependence of contrast decay time $\tau$. The strong dependence on density provides evidence that the contrast decay is not due to single-particle effects, but rather due to dipole-dipole interactions. The fit line shows a $1/N$ dependence on particle number, and density. (\textbf{b})~The fit-determined oscillation frequency $f$ is shown as a function of total molecule number. The fit frequency is roughly constant for different molecule densities, with an average value of 48~Hz being in excellent agreement with the predicted largest interaction energy of $|0,0\rangle$ and $|1,-1\rangle$ molecules in our system (ab initio prediction of 52~Hz).
}
\end{figure}

The ability to systematically study the Ramsey fringe contrast dynamics across many densities provides us with a powerful tool. For a fixed shape to the particle distribution, any loss of contrast due to single-particle effects should be independent of molecule number. The influence of dipolar interactions, however, should fundamentally depend on the density and spacing of the lattice-trapped molecules. Shown in Fig.~\ref{fig:density}~(a) are the fit \emph{coherence times} of the Ramsey fringe contrast dynamics, systematically studied over a large range of molecule numbers (densities). To quantify the rate of contrast decay, here we simply fit to an empirical function $Ae^{-T/\tau}+B\cos^2(\pi f T)$ that features coherent oscillations (at frequency $f$) on top of an overall exponential decay over a coherence time $\tau$~\footnote{This empirical fitting function is known to systematically deviate from the data - the fringe contrast decays quadratically at short times~\cite{yan2013:dipole-dipole} and oscillates at multiple frequencies. Still, this simple single-frequency fit reliably captures the most salient features of the contrast dynamics. Moreover, the fit-determined \emph{coherence time} is robust to the chosen form of the fitting function - similar behavior is found for a purely exponential fitting function $e^{-T/\tau}$, as used in Ref.~\cite{Hazzard2014}.}, with empirical fits shown along with the data in Fig.~\ref{fig:scheme}~(d). The fit-determined coherence times $\tau$ and oscillation frequencies $f$ are shown in Fig.~\ref{fig:density}~(a,b) across a range of particle densities (with several datasets excluded from the frequency determination, due to an insufficient sampling that does not fulfill the Nyquist sampling criterion). With densities varied by almost an order of magnitude, a strong dependence of the coherence time is seen, along with almost no change to the observed oscillation frequency.

The contrast decay times $\tau$ in Fig.~\ref{fig:density}~(a) roughly follow a $1/N$ dependence to the particle numbers (densities). This is in agreement with theory estimates for dipolar interactions, which scale as $1/r^3$. Because the average interparticle spacing in our three-dimensional sample of molecules scales as $\langle r \rangle \propto N^{-1/3}$, the average - and typical \emph{spread} in - dipolar interaction energy is proportional to $N$. The resultant coherence time $\tau$ (inverse to the \emph{decoherence rate}) scales as $1/N$. The fit-determined oscillation frequencies $f$ are shown in Fig.~\ref{fig:density}~(b), exhibiting little variation with the total molecule number $N$. The average value of 48(2)~Hz is in close agreement with the predicted largest nearest-neighbor interaction energy of $|J_\perp/h|=52$~Hz. This ab initio prediction for the $|0,0\rangle$ and $|1,-1\rangle$ rotational states at a spacing of 532~nm takes into account a small reduction of the effective dipole moment due to a mixing of different hyperfine states at the level of a few percent~\cite{yan2013:dipole-dipole}. Whereas the spread in the average interaction energy scales strongly with density, this largest pairwise configuration energy is fixed to a discrete value by the underlying lattice, and is thus mostly independent of molecule density. This largest energy scale shows up most strongly in the data, as it is furthest separated from other configuration energies $J_\perp V_{dd}$ and most easily observed on the short experimental timescales, however we will later present evidence for multiple discrete frequencies relating to different molecular configurations in our system.

\subsubsection{Reversing pairwise entanglement through a multi-pulse echo sequence}

\begin{figure}[b]
\includegraphics[height=6cm]{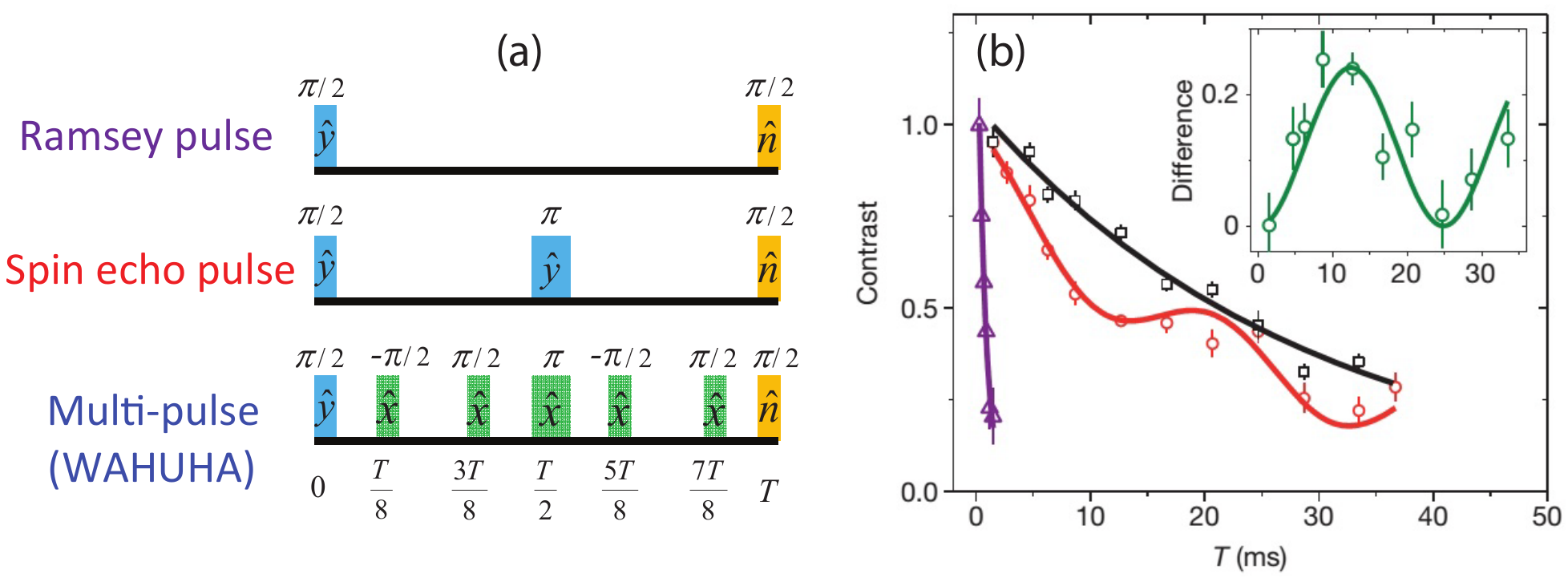}
\caption{\label{fig:decoupling}
Spin exchange oscillation and dipole-dipole decoupling (from Ref.~\cite{yan2013:dipole-dipole}).
(\textbf{a})~Ramsey pulse sequences used for studying dipolar interactions. Shown, from top to bottom, are a normal Ramsey sequence, a Ramsey spin echo sequence that mitigates single-particle dephasing, and a multi-pulse WAHUHA sequence that reverses effects of both single-particle dephasing and pairwise dipole-dipole interactions.
(\textbf{b})~The Ramsey fringe contrast decay as a function of time under the three different pulse sequences. For the two-pulse Ramsey sequence, global inhomogeneities of the effective ``magnetic field'' leads to Ramsey coherence decay over just 1~ms (triangles). The spin echo pulse effectively removes contrast loss due to single-particle dephasing, so that spin-exchange interactions play the dominant role in the contrast decay (circles). The multi-pulse WAHUHA sequence acts as an ``interaction echo'' for pairs of molecules, suppressing contrast oscillations and extending the coherence decay time (squares). \textbf{Inset}~The difference in contrast between the the multi-pulse WAHUHA sequence and the spin echo case, demonstrating clear oscillations.
}
\end{figure}

Because of the diluteness of our molecular samples, we expect that the contrast dynamics studied in Fig.~\ref{fig:scheme} and Fig.~\ref{fig:density} are largely dominated by interactions between isolated pairs of molecules. For a fully isolated pair of molecules, having a spin-exchange coupling $J_{ij} = J_\perp V_{dd}^{ij}$, the Ramsey fringe contrast will undergo coherent oscillations as the molecules undergo entanglement oscillations. Explicitly, just after the initial $(\pi/2)_y$ excitation pulse, the two molecules are in a product state $|$$\rightarrow$$\rightarrow$$\rangle = (1/\sqrt{2})(|$$\downarrow$$\rangle+|$$\uparrow$$\rangle) \otimes (1/\sqrt{2})(|$$\downarrow$$\rangle+|$$\uparrow$$\rangle)$, which can be rexpressed as $(|$$\downarrow$$\downarrow$$\rangle+|$$\uparrow$$\uparrow$$\rangle+|$$\downarrow$$\uparrow$$\rangle+|$$\uparrow$$\downarrow$$\rangle)/2$. This combination of triplet configurations (reflecting the symmetry of the initial $|$$\downarrow$$\downarrow$$\rangle$ state that is preserved during the excitation pulse) can be divided into different parity sectors. The even parity portion, $(|$$\downarrow$$\downarrow$$\rangle+|$$\uparrow$$\uparrow$$\rangle)$, is unaffected by the purely exchange interaction Hamiltonian $\hat{H}^{\mathrm{int}}$. That is, these two configurations are zero energy eigenstates of $\hat{H}^{\mathrm{int}}$. In contrast, the odd parity triplet configuration $(|$$\downarrow$$\uparrow$$\rangle+|$$\uparrow$$\downarrow$$\rangle)$ acquires an interaction phase shift $e^{-i(J_{ij}T/\hbar)}$ during an evolution time $T$. After a time $T = h/(4J_{ij})$ the two molecules have become maximally entangled, and may no longer be described as a simple product state because of their strong correlations. Measurement of the spin coherence at this time will reveal zero Ramsey fringe contrast, while the system will later return to a coherent product state at time $T = h/(2J_{ij})$, and continue undergoing entanglement oscillations as time goes on. For many different configuration energies $J_{ij}$, a simple sum over coherence oscillations of different pairwise configurations will result in an overall decay in the Ramsey fringe contrast (quadratic at short times as the oscillations are all initially in phase~\cite{yan2013:dipole-dipole}).

Using the powerful set of tools developed for the study of spins in NMR experiments, we may hope to directly probe this pairwise entanglement process through application of a tailored multi-pulse echo sequence. In general, multi-pulse echo sequences can be used to handle dynamical (non-static) field inhomogeneities, through so-called dynamical decoupling~\cite{Sagi-DD-noise,DynDec-Noise,Bishof-DD-specanal}, to mitigate dephasing and extend coherence times~\cite{du,bollinger,maurer:room-temperature_2012}. For our interaction Hamiltonian, however, a complete reversal of all pairwise correlations (leading to loss of Ramsey fringe contrast) can be achieved at all evolution times. The required pulse sequence for achieving this remarkable feat is known from NMR studies of dipolar systems~\cite{waugh:approach_1968}, and we refer to it as the WAHUHA pulse sequence. As shown in Fig.~\ref{fig:decoupling}~(a), this special sequence mainly differs from the normal Ramsey spin echo pulse sequence by the application of additional $\pi/2$ pulses that are phase-shifted by 90 degrees with respect to the initial excitation pulse (i.e. involving rotations about the $\hat{x}$ spin axis). In much the same way that a spin echo $\pi$ pulse can rephase single spins in an inhomogeneous field, these $(\pi/2)_x$ pulses act to rephase interacting pairs with different (inhomogeneous) interaction energies. The spin echo $\pi$ pulse swaps population between the two eigenstates of the local field Hamiltonian $h_i \hat{S}^z_i$, such that over one symmetric cycle (with an echo pulse applied during the middle of the evolution) differential phase accumulation is completely cancelled out. In analogy to this, the $(\pi/2)_x$ pulse acts as an ``interaction echo'' for molecule pairs under the exchange interaction $\hat{H}^{\mathrm{int}}$. The $(\pi/2)_x$ pulse perfectly swaps population between the even and odd triplet configurations~\cite{Scha-QM}, i.e. $(|$$\downarrow$$\downarrow$$\rangle+|$$\uparrow$$\uparrow$$\rangle) \rightleftarrows (|$$\downarrow$$\uparrow$$\rangle+|$$\uparrow$$\downarrow$$\rangle)$. Thus, by simply applying a single $(\pi/2)_x$ pulse halfway throughout the evolution time, correlations built up due to the dipolar exchange interaction will be ``rephased'' for all of the different pairwise configuration energies $J_{ij}$ in our system. The many additional pulses that are used in the multi-pulse sequence (central $(\pi)_x$ pulse and several $(\pm\pi/2)_x$ pulses), as shown in Fig.~\ref{fig:decoupling}~(a), are to preserve the benefits of the normal spin echo protocol that removes single-particle dephasing.

Figure~\ref{fig:decoupling}~(b) summarizes the Ramsey contrast decay for three different pulse sequences. With a simple two-pulse Ramsey sequence (no spin echo pulse), the contrast decay time is less than 1~ms, mainly caused by single particle dephasing. With the addition of a single spin echo pulse, the contrast decay time can be extended to $\sim$20~ms (and up to $\sim$80~ms for low molecule densities), limited by dipolar interactions. In addition to the decay of Ramsey fringe contrast, clear oscillations can also be observed due to dipolar exchange interactions. When we apply the multi-pulse WAHUHA sequence, oscillations in the contrast are suppressed, and the overall coherence timescale is slightly increased. This pulse sequence reverses the build-up of correlations due to pairwise dipolar interactions. The remaining contrast dynamics can be attributed in part to deviations from the oversimplified picture of pairwise interactions, i.e. any configuration of three or more interacting molecules where this simple scheme fails. Additionally, technical imperfections (such as errors in the single-pulse fidelity compounded over many pulses) or dynamical ``field'' noise may contribute to the decay at long times.

\subsubsection{Controlling dipolar interactions through choice of rotational states}

\begin{figure}
\includegraphics[height=5cm]{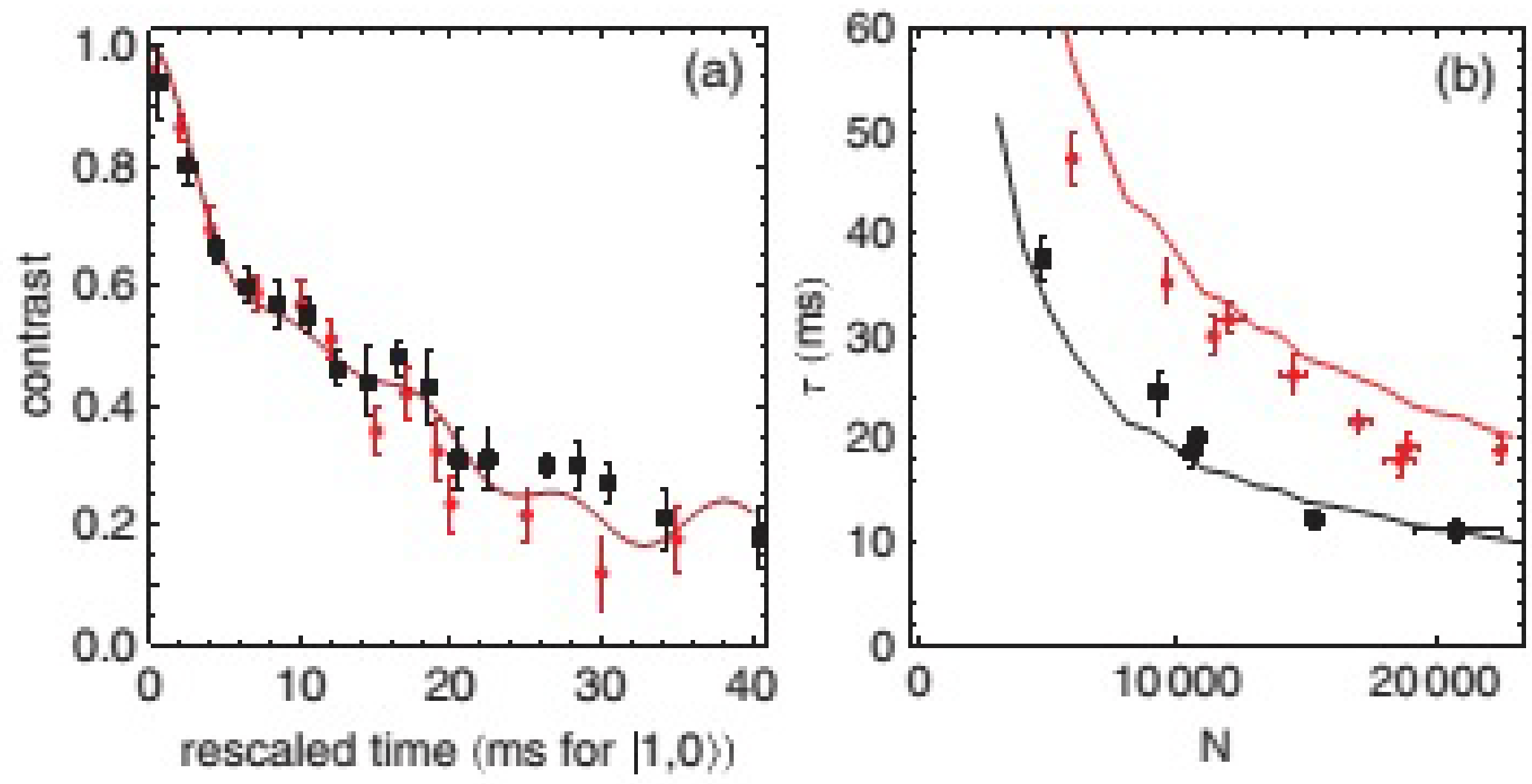}
\caption{\label{fig:diffs}
Control of dipole-dipole interactions through rotational state choice (from Ref.~\cite{Hazzard2014}). (\textbf{a})~Contrast decay for two choices of excited rotational state, for either $|$$\uparrow$$\rangle = |1,-1\rangle$ (red circles) or $|1,0\rangle$ (black squares). For $|1,-1\rangle$, we rescale the time axis by factor of 1/2. The nearly complete overlap of the two sets of data confirms the expected factor of 2 difference in dipolar interaction strength, and further suggests that essentially only dipole-dipole interactions influence the spin coherence dynamics. (\textbf{b})~Contrast decay time vs. molecule number (density). Decreased coherence times with higher densities, consistent with a $1/N$ dependence on molecule number, are seen for both data sets. For the case $|$$\uparrow$$\rangle =|1,0\rangle$, the coherence decay times are roughly a factor of 2 faster than those for $|$$\uparrow$$\rangle = |1,-1\rangle$, consistent with the stronger dipolar exchange interaction. The solid lines in both (a) and (b) are theoretical predictions based on Eq.~\ref{eqn:hamiltonianA} with only a single global fitting parameter (the proportionality factor relating molecule number to peak lattice filling fraction), as detailed in Ref.~\cite{Hazzard2014}.
}
\end{figure}

One of the great experimental features of polar molecules is that their resonant dipole moment may be widely tuned by application of a dc electric field. Even at zero electric field, we may discretely vary the transition dipole moment of our polar molecules, and thus the strength of excitation exchange coupling $J_\perp$, by the choice of rotational excited state. As a cursory inspection of Fig.~\ref{fig:polarizability}~(a) reveals, ``magic'' conditions can also be chosen to minimize the differential ac Stark shift (or more precisely, to minimize spatial variations of the differential ac Stark shift) between the states $|0,0\rangle$ and $|1,0\rangle$. As motivated earlier, use of $|1,0\rangle$ as the excited state $|$$\uparrow$$\rangle$ leads to a factor of 2 enhancement in $J_\perp$ as compared to the states $|1,\pm 1\rangle$. By simply changing the frequency of the applied microwaves to be resonant with the $|0,0\rangle$ to $|1,0\rangle$ transition, and with the slight modification of lattice laser polarizations to minimize ``field'' inhomogeneities, we can directly compare the dipolar interaction-driven contrast dynamics for these two different excited rotational states. Figure~\ref{fig:diffs} shows a summary of this comparison. In Fig.~\ref{fig:diffs}~(a), we show the contrast dynamics for nearly identical molecule numbers (densities), but with a different choice of the rotational excited state. Here, the time axis for the $|1,-1\rangle$ state is scaled by a factor of 1/2, and we find a nearly complete collapse of the two data sets onto one another. This agreement confirms the expected difference in dipolar interaction energy for these two cases. Moreover, the lack of any substantial deviations suggests that the contrast dynamics are driven almost solely by the coherent dipolar interactions. This agreement between the two sets is seen more fully by again varying the molecule density over a large range. As shown in Fig.~\ref{fig:diffs}~(b), we again see the expected $1/N$ dependence to the contrast decay time, as well as a roughly factor of two difference in the contrast decay times when comparing the two excited spin states. As discussed in Ref.~\cite{Hazzard2014}, the solid lines in (a) and (b) are theoretical predictions based on Eq.~\ref{eqn:hamiltonianA}, determined by numerical simulations using the ``moving average cluster expansion'' (MACE) method, fixed by only a single global fitting parameter (the proportionality factor relating molecule number to lattice filling fraction).

\subsubsection{Evidence for multiple interaction energies}

Finally, we discuss the observation of multiple interaction energies in the Ramsey fringe contrast dynamics. As motivated in Fig.~\ref{Fig:geometric_factor}~(c), many different exchange interaction energies should be present in a system of interacting dipolar particles. However, for the data presented in Fig.~\ref{fig:scheme}~(d) and Fig.~\ref{fig:density}~(b), only the largest interaction energy ($|J_\perp|$) was seen directly in the contrast dynamics for the $|1,-1\rangle$ excited state. While there was strong evidence for the presence of many different frequencies - i.e. in the appearance of an overall contrast decay, the success of the WAHUHA multi-pulse sequence, and the excellent agreement with the theory curves in Fig.~\ref{fig:diffs} (which sample all possible interaction configurations) - it is reasonable to question why only a single frequency was clearly resolved. For one, the observed frequency of $|J_\perp/h| \sim$50~Hz relates to the largest pairwise interaction energy in our system (also the most well-separated from other energies). Related to this, the duration (40~ms) over which and the rate (1 point every 2-4~ms) at which the contrast dynamics data were sampled effectively served to filter out lower frequencies.

\begin{figure}
\includegraphics[height=8cm]{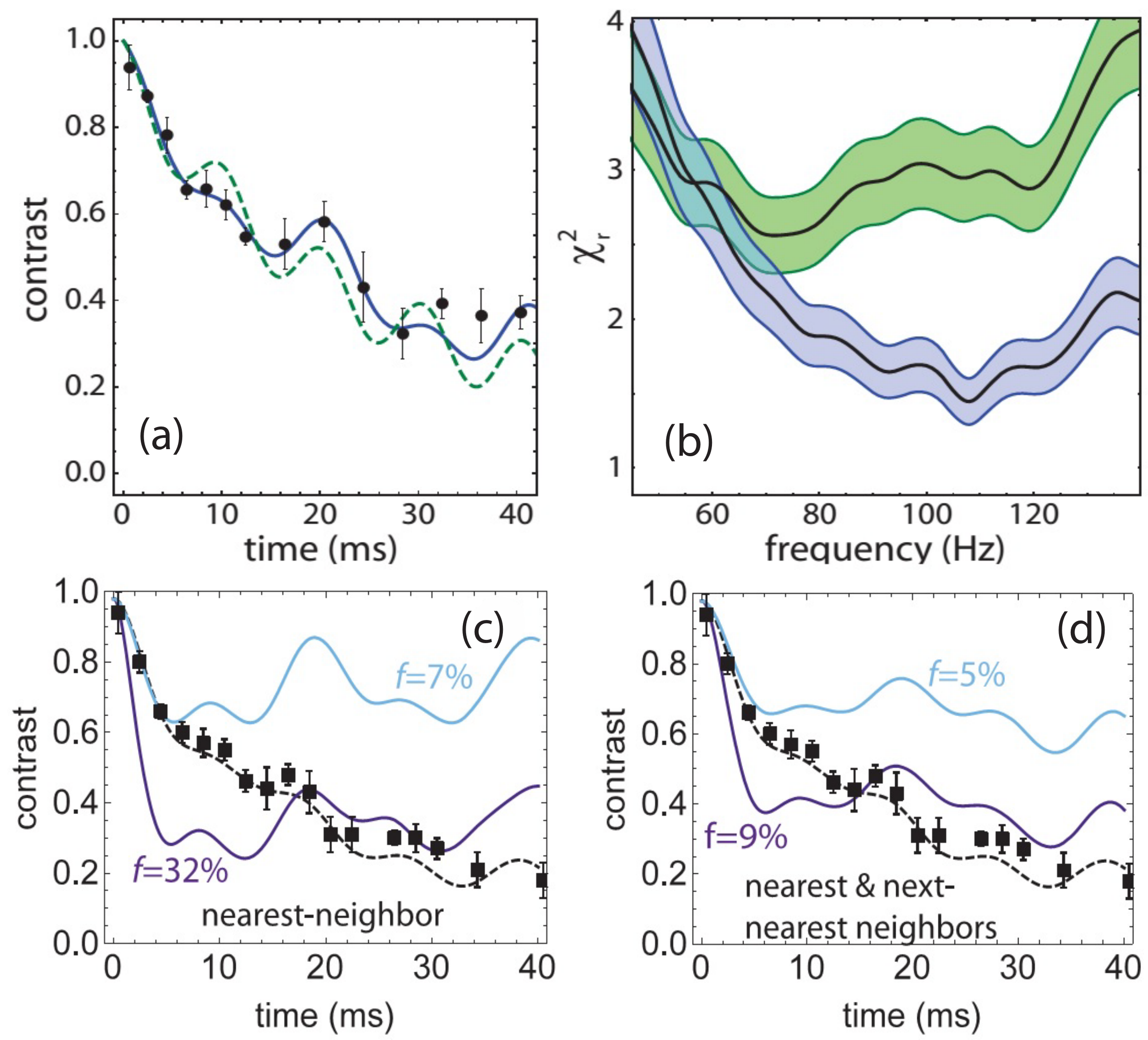}
\caption{\label{fig:multi_frequencies}
Evidence for multiple and long-ranged interaction energies (from Ref.~\cite{Hazzard2014}).
(\textbf{a})~Comparison of the contrast dynamics data (for $|$$\uparrow$$\rangle =|1,0\rangle$), with empirical fits featuring a single oscillation frequency (green dashed line) and three frequencies (blue solid line), as described in the text. (\textbf{b})~Cumulative reduced chi-squared ($\chi^2_r$) of multiple data sets for both the single-frequency fitting (green) and three-frequency fitting (blue), as a function of the primary fit frequency.
(\textbf{c,d})~Theoretically calculated contrast dynamics for (c) nearest-neighbor interactions and (d) interactions to next-nearest-neighbors for two fillings (indicated). These finite-range theory curves are compared with the contrast dynamics data for roughly 12,000 molecules ($|$$\uparrow$$\rangle =|1,0\rangle$) and a full theoretical prediction (dashed line).
}
\end{figure}

By looking at the contrast dynamics for the case of $|$$\uparrow$$\rangle =|1,0\rangle$, which plays host to stronger interactions, we can perhaps hope to resolve multiple frequency components. Looking at Fig.~\ref{Fig:geometric_factor}~(c), we find that the three largest pairwise interaction energies for our lattice geometry and quantization axis should be at $|J_\perp|$, $J_\perp / \sqrt{2}$, and $J_\perp/2$ (we note that if the quantization axis was aligned along a lattice axis, interaction energies of $|2 J_\perp|$ would also be present). Figure~\ref{fig:multi_frequencies}~(a) shows the contrast dynamics for the case of $|$$\uparrow$$\rangle =|1,0\rangle$, fit with an empirical fitting function of the form
\begin{equation}
A\cos^2(\pi f T)+B\cos^2(\frac{\pi f T}{\sqrt{2}})+C\cos^2(\frac{\pi f T}{2})+(1-A-B-C) e^{-T/\tau}
\end{equation}
containing three oscillation frequencies fixed by the expected ratios (solid blue line), as well as a single-frequency fit (dashed green). For this exemplary data set, there is a much better apparent agreement with the multifrequency fit. To be more rigorous, we can also examine the reduced chi-squared of the two different fitting functions, which accounts for the additional fit parameters of the multi-frequency fit. Fig.~\ref{fig:multi_frequencies}~(b) shows the results of a cumulative (across all available data sets with sufficient sampling rates) reduced chi-squared analysis for the single-frequency (green) and multi-frequency (blue) functional forms. This analysis suggests a (statistically significant) better agreement with the presence of multiple frequencies, and moreover determines the most likely value of $J_\perp/h$ to be 108~Hz, which is close to the ab initio theoretical prediction of 104~Hz.

For both of these empirical fitting functions, an exponentially decaying contribution was included, to qualitatively capture the loss of Ramsey fringe contrast due to the expected broad distribution of interaction energies at low energy. Stronger evidence for multiple interaction energies may be gleaned from the data by asking how the dynamics would appear if only one (or only a few) interaction energies were present. This question was addressed in Ref.~\cite{Hazzard2014}, and Figs.~\ref{fig:multi_frequencies}~(c,d) compare contrast dynamics data to several different theory curves that include interactions up to different interparticle distances. As in Fig.~\ref{fig:diffs}, a full numerical simulation based on solution by the MACE method is included as a solid dashed line in Figs.~\ref{fig:multi_frequencies}~(c,d). In Fig.~\ref{fig:multi_frequencies}~(c), the solid lines show example theoretical predictions for the contrast dynamics if only nearest-neighbor configurations are included. Here, when the short-time dynamics - which should be dominated by nearest neighbors - are fit well by the theoretical curve, the long-time decay is not captured at all. Figure~\ref{fig:multi_frequencies}~(d) expands on this by including also next-nearest-neighbor configurations. While the appearance of several extra configuration energies helps somewhat with the agreement at long times, the finite-range theory predictions still deviate significantly from the experimental data. Thus, while not individually resolved, the decay of spin coherence dynamics at long times provides perhaps the strongest evidence for the presence of many different interaction energies in our lattice trapped ensemble of polar molecules.
\section{Conclusions and future prospects}
\label{sec:sec4outtro}

Systems of ultracold polar molecules are poised to serve as an ideal platform for the study of strongly correlated many-body physics. The JILA experiments on rotational excitation exchange of lattice-trapped KRb molecules~\cite{yan2013:dipole-dipole,Hazzard2014} have shown that these systems support strong and long-ranged spin-spin interactions, which are in excellent agreement with theory predictions based on known dipolar couplings. Through comparison with numerical simulations~\cite{Hazzard2014}, there is added evidence that this system realizes a quantum spin-1/2 XY Hamiltonian, with long-range dipolar couplings. Importantly, the molecule lifetimes can be very long in these experiments~\cite{chotia:long-lived_2012}, and there is as of yet scant evidence for any sources of appreciable decoherence~\cite{yan2013:dipole-dipole,Hazzard2014}. Recent advances in the synthesis of molecules from a quantum gas mixture have additionally allowed for the achievement of very low entropy molecular samples with high lattice filling~\cite{Moses-LowEntropy}. This unique combination of low entropies, long coherence times, and strong non-local interactions makes these systems well-suited to study the dynamics of quantum correlations and entanglement. The correlated dynamics of particle and spin transport in ultracold dipolar matter seems especially promising for the study of emergent behavior in frustrated systems~\cite{Balents-SpinLiquid,Tewari-EmergenceDipolar}.

One of the greatest sources of excitement in the area of ultracold molecule research stems from the large number of research groups that have joined and are joining the effort. Particularly encouraging are the recent achievements of cold and dense ground state molecular gases of RbCs at Innsbruck~\cite{Takekoshi:STIRAP2014} and Durham~\cite{Molony:STIRAP2014}, NaK at MIT~\cite{Park-NaK-GS-2015}, and NaRb at Hong Kong~\cite{NaRbGas}. There are now a handful of groups ready to experiment with ultracold ground state molecules, with a number of other groups and molecular species on their way. The diversity of molecular species is growing, as there are now bosonic and fermionic molecular gases, chemically stable and unstable molecules, and a larger range of available electric dipole moments, with many other important differences still to be found out. Many non-bi-alkali molecules formed from quantum gas mixtures, such as those involving alkaline earth atoms~\cite{Ivanov-LiYb,Pasq-RbSr,Dowd-LiYb}, lanthanide atoms, or chromium, will likely be joining the fold soon as well. Continued advances in the direct cooling and trapping of molecules are likely to open up the quantum regime to a still wider assortment of molecular species.

In addition to the inherent diversity of different molecular species, added experiments will also bring a plurality of tools and techniques for manipulating and probing molecular samples. The application of high-resolution imaging, such as through quantum gas microscopy~\cite{Bakr-Science-2010,Sherson-nature-2011}, to molecular samples is extremely exciting as it will provide unique opportunities for the study of entanglement and quantum correlations in a long-range interacting system. The rich internal structure of molecules, or more specifically the large number of internal states that may be coupled through optical, microwave, and radiofrequency fields, promises to allow for a unique control over the range and nature of dipolar interactions via state dressing~\cite{Buchler-Crystal-2007,Micheli-tailoring-2007,manmana:topological_2013,Yao2013:Chern,Wall-SymmTop}. Other tools and techniques, such as the trapping of molecules in novel lattice geometries, the development of spin-dependent lattice potentials, and the engineering of artificial classical gauge fields for molecules, should also open many new avenues of research.

Lastly, we remark that as the capability to study strongly interacting systems of polar molecules progresses, researchers will no doubt continue to be pushed and challenged by those working on related AMO systems. Trapped ion simulators~\cite{Kim-Frustr,Islam-Onset-2011,islam:emergence_2013,Richerme-Prop-2014,Blatt-Prop-2014}, lattice trapped alkali atoms with tunneling-mediated exchange interactions~\cite{trotzky:time-resolved_2008,greif:quantum_2012,fukuhara:quantum_2013,Hart-Hulet-2014,FourBody}, and systems of individually-trapped Rydberg atoms~\cite{Browaeys-DipDip-14,Browaeys-ExcitationTransfer-15} as well as Rydberg-dressed atoms~\cite{RydDressLattice} have all been shown to be well-suited to the study of quantum magnetism and interacting spin systems. An even more direct competition will likely continue to come from systems of magnetically dipolar atomic gases, where the past few years have seen remarkable advances in the study of nonequilibrium quantum magnetism~\cite{dePaz-Chrom-QM-2013,bruno-nonequ-2-2015} and long-range interactions in itinerant systems~\cite{Ferlaino-Extended-2015,Pfau-Crystal-2015}.

\section*{Acknowledgements}
We acknowledge helpful discussions with the JILA KRb team led by Jun Ye and Deborah Jin, especially Steven Moses and Jacob Covey, as well as with Goulven Qu\'{e}m\'{e}ner, Ana Maria Rey, Kaden Hazzard, Michael Wall, Johannes Schachenmayer, and Bihui Zhu.
%
\bibliographystyle{iopart-num}
\bibliography{BibMolRev}
%
%

\end{document}